\documentclass[
    twocolumn,
    superscriptaddress,
    amsmath,amstex,amssymb,
    aps, pra,
    floatfix,
    longbibliography,
    reprint]{revtex4-2}

\usepackage{graphicx}
\usepackage{dcolumn}%
\usepackage{bm}
\usepackage{lipsum}
\usepackage{gensymb}
\usepackage{physics}
\usepackage{xcolor}

\usepackage{hyperref}
\hypersetup{
  colorlinks   = true, 
  urlcolor     = blue, 
  linkcolor    = red, 
  citecolor   = red 
}

\begin{document}

\title{Towards a hybrid 3D transmon qubit with\\topological insulator-based Josephson junctions}

\author{Sheng-Wen~Huang}
    \affiliation{%
     Department of Physics and Astronomy, Purdue University, West Lafayette, IN 47907, USA}%
\author{Ramya~Suresh}
    \affiliation{%
     Department of Physics and Astronomy, Purdue University, West Lafayette, IN 47907, USA}%
\author{Jian~Liao}
    \affiliation{%
     Department of Physics and Astronomy, Purdue University, West Lafayette, IN 47907, USA}%
\author{Botao~Du}
    \affiliation{%
     Department of Physics and Astronomy, Purdue University, West Lafayette, IN 47907, USA}%
\author{Zachary~Miles}
    \affiliation{%
     Department of Physics and Astronomy, Purdue University, West Lafayette, IN 47907, USA}%
\author{Leonid~P.~Rokhinson}
    \affiliation{%
     Department of Physics and Astronomy, Purdue University, West Lafayette, IN 47907, USA}%
     \affiliation{Elmore Family School of Electrical and Computer Engineering, Purdue University, West Lafayette, Indiana 47907, USA}
     \affiliation{Birck Nanotechnology Center, and Purdue Quantum Science and Engineering Institute,
    Purdue University, West Lafayette, Indiana 47907, USA}
\author{Yong~P.~Chen}
    \affiliation{%
     Department of Physics and Astronomy, Purdue University, West Lafayette, IN 47907, USA}%
     \affiliation{Elmore Family School of Electrical and Computer Engineering, Purdue University, West Lafayette, Indiana 47907, USA}
     \affiliation{Birck Nanotechnology Center, and Purdue Quantum Science and Engineering Institute,
    Purdue University, West Lafayette, Indiana 47907, USA}
\author{Ruichao~Ma}
    \email{maruichao@purdue.edu}
    \affiliation{%
     Department of Physics and Astronomy, Purdue University, West Lafayette, IN 47907, USA}%

\date{\today}

\begin{abstract}
Superconducting quantum circuits provide a versatile platform for studying quantum materials by leveraging precise microwave control and utilizing the tools of circuit quantum electrodynamics (QED). Hybrid circuit devices incorporating novel quantum materials could also lead to new qubit functionalities, such as gate tunability and noise resilience. Here, we report experimental progress towards a transmon-like qubit made with a superconductor-topological insulator-superconductor (S-TI-S) Josephson junction using exfoliated $\mathrm{BiSbTeSe_2}$. We present a design that enables us to systematically characterize the hybrid device, from DC transport of the S-TI-S junction, to RF spectroscopy, to full circuit QED control and measurement of the hybrid qubit. In addition, we utilize a high-quality-factor superconducting cavity to characterize material and fabrication-induced losses, thereby guiding our efforts to improve device quality. 

\begin{description}
    \item[Keywords]
 Superconducting circuits, topological insulator, qubit design 
\end{description}

\end{abstract}

\maketitle

\section{Introduction\label{sec:intro}}

Superconducting (SC) quantum circuits are a leading platform for quantum information science, offering flexibility and scalability for quantum computation, simulation, and sensing \cite{Kjaergaard2020-rt, Blais2020-ef, Carusotto2020-ct}. Their rapid development has been driven by advances in circuit design and architecture, improved coherence through materials and fabrication optimization, and advancements in control and readout capabilities via hardware and software development.

There is a growing effort to integrate novel materials into SC circuits, aiming to enhance qubit performance and explore new functionalities. For example, hexagonal boron nitride has been employed as a low-loss dielectric to reduce device footprint \cite{Antony2021-tf, Wang2022-qn}, and may provide a low critical-current-noise alternative to conventional aluminum oxide-based Josephson junctions \cite{Tian2021-mu}. Superconductor–semiconductor Josephson junctions with gate tunability have been realized using nanowires \cite{Larsen2015-hu, de-Lange2015-de}, 2D electron gases \cite{Casparis2018-ar}, and graphene \cite{Kroll2018-ix, Wang2019-il, Chiu2025-or}.

Conversely, SC circuits can also serve as sensitive probes of quantum materials, leveraging their strong electromagnetic coupling with other physical degrees of freedom in hybrid quantum systems \cite{Clerk2020-io} and the precise control
and readout of single microwave excitations using the toolset of circuit quantum electrodynamics (circuit QED) \cite{Blais2021-bn}. Superconducting resonator circuits have been used to measure microwave losses in dielectric substrates \cite{Read2023-qy, Checchin2022-ek}, probe and control magnon dynamics \cite{Tabuchi2014-ij}, and study unconventional superconductivity \cite{Phan2022-py, Bottcher2024-kc}. This bidirectional interplay between SC circuits and novel materials forms a co-design cycle: SC circuits enable characterization of novel materials, while insights from these studies inform the development of improved qubits and hybrid devices.

Among various quantum materials, topological insulators (TIs) have attracted significant interest \cite{Hasan2010-id, Qi2011-ua}. TIs can host protected surface states that are robust against local perturbations. When proximitized by a superconductor, TIs can exhibit induced topological superconductivity and support exotic excitations with potential applications in topological quantum computing \cite{FuKane2008-bq, Schrade2018-aa}. These novel properties have motivated proposals to use circuit QED techniques to probe and manipulate such excitations in TI-based devices \cite{Hassler2011-zr, van-Heck2015-en, Ren2024-ae}. 

In this work, we present the design and initial experimental results for a hybrid 3D transmon qubit incorporating a superconductor–topological insulator–superconductor (S–TI–S) Josephson junction using exfoliated $\mathrm{BiSbTeSe_2}$, a three-dimensional topological insulator. Previously, TI-transmon qubits using planar 2D designs have been realized using S–TI–S junctions fabricated from $\mathrm{(Bi,Sb)_2Te_3}$ nanoribbons \cite{Schmitt2022-il} and $\mathrm{Bi_2Se_3}$ nanowires \cite{Sun2023-rj}.
Our 3D TI-transmon design enables the use of the same device for DC transport, RF spectroscopy, and time-domain circuit QED measurements. This feature is crucial for the systematic characterization of these hybrid devices. In addition, we utilize high-quality-factor superconducting cavities to characterize material- and fabrication-induced losses, thereby guiding our fabrication and design optimization.

\begin{figure*}[!ht]
    \centering
    \includegraphics[width=0.8\linewidth]{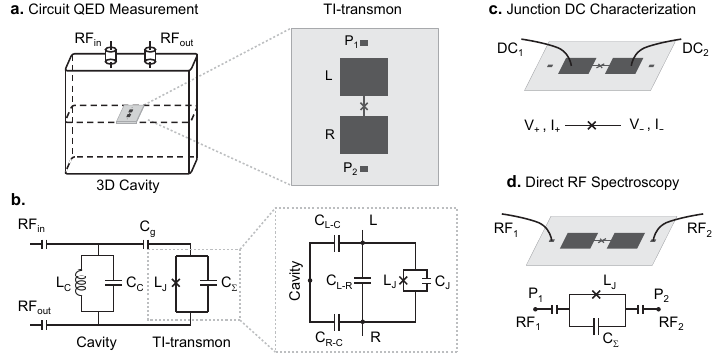}
    \caption{Overview of the device design and measurement scheme. (a) Illustration of the circuit QED measurement setup with a zoomed-in view of the TI-transmon qubit. (b) Effective circuit diagram for (a), the zoom-in shows the equivalent circuit of the 3D transmon. (c) DC transport characterization of the S-TI-S junction. (d) Direct RF spectroscopy of the TI-transmon via the coupling pads $\text{P}_\text{1,2}$. Measurements in (c) and (d) can be performed in either the 3D cavity or a 2D PCB-based package. Geometries are for illustrative purposes and not drawn to scale.}
    \label{fig:1-design}
\end{figure*}

The remainder of this paper is organized as follows. Section~\ref{sec:design} details the TI-transmon design, including the S–TI–S junction and integration into a 3D circuit QED architecture. Section~\ref{sec:fab} outlines the fabrication process. Section~\ref{sec:results} presents measurement setups for DC and microwave characterization. Section~\ref{sec:conclusion} concludes with a discussion of future directions.

\section{Device Design\label{sec:design}}

\subsection{BSTS-based TI-transmon}

The transmon qubit is a widely adopted superconducting (SC) qubit architecture, effectively realized as a nonlinear microwave LC resonator comprising a linear capacitor and a Josephson junction acting as a nonlinear inductor \cite{Koch2007-vs, Roth2023-lq}. With typical transition frequencies in the $4\text{–}10$\,GHz microwave range, transmons are operated at millikelvin temperatures, where thermal excitation is strongly suppressed \cite{Gao2021-kj}.

In the TI-transmon, we replace the conventional $\mathrm{Al/AlO_x/Al}$ junction with an S-TI-S Josephson junction based on the 3D TI material $\mathrm{BiSbTeSe_2}$ (BSTS). BSTS has previously shown robust topological surface state conduction with minimal bulk contribution, as evidenced by quantum Hall measurements \cite{Xu2014-fv}. Prior studies of S-BSTS-S junctions have demonstrated gate-tunable supercurrents and multiple Andreev reflections \cite{Jauregui2018-fr, Kayyalha2019-xc}, as well as skewed, non-sinusoidal current-phase relations consistent with ballistic transport through surface states \cite{Kayyalha2020-vk}.

We compare our BSTS-based junction design with earlier TI-transmon implementations. In Ref.\,\cite{Schmitt2022-il}, the TI used is $\mathrm{(Bi_{0.06}Sb_{0.94})_2Te_3}$ grown via area-selective molecular beam epitaxy. The devices incorporate roughly 100\,nm wide and 20\,nm thick TI nanoribbons and niobium SC contacts formed by stencil lithography, with measured energy relaxation times $T_1\approx30$\,ns. In Ref.\,\cite{Sun2023-rj}, $\mathrm{Bi_2Se_3}$ nanowires synthesized by a vapor-liquid-solid method were integrated with pre-patterned SC circuits, achieving $T_1\approx500$\,ns using Ti/Al contacts. In both cases, the dephasing times $T_2$ of the TI-transmons were below 10\,ns.
Here, we utilize exfoliated BSTS flakes with lateral dimensions of several microns, which are significantly larger than those of the nanoribbon and nanowire geometries in previous works. The use of BSTS as a 3D intrinsic TI might offer high coherence junctions with lower dissipation due to its highly insulating bulk \cite{Xu2014-fv}.

We have explored several SC contact materials for our S-BSTS-S junctions. These include Niobium contacts following previous studies \cite{Kayyalha2019-xc, Kayyalha2020-vk, Schmitt2022-il}, Molybdenum-Rhenium (MoRe), and aluminum contacts that incorporate a thin titanium interlayer \cite{Sun2023-rj}. 
Additional interlayer materials have been investigated to optimize S-TI-S junction quality \cite{Jalil2024-cc}.

\subsection{3D circuit QED for TI-transmon}

To realize a hybrid TI-transmon qubit, we employ a 3D circuit QED architecture \cite{Paik2011-fm, Krasnok2024-xb} rather than the planar geometries used in prior work \cite{Schmitt2022-il, Sun2023-rj}. The 3D design simplifies fabrication and enables both DC transport and microwave characterization in a unified package.

As illustrated in Fig.\,\ref{fig:1-design}a, the qubit design consists of two SC islands forming a split-capacitor geometry, connected by a S-TI-S junction. When operated as a qubit, the device is housed inside a rectangular 3D microwave cavity, which provides an engineered electromagnetic environment and serves as the readout resonator for the qubit.

We target a junction critical current of $I_c \approx 30$\,nA for the S-TI-S junction, corresponding to a Josephson energy $E_J/\hbar = I_c / 2e \approx 2\pi \times 14.9$\,GHz. Here $\hbar$ is the reduced Planck constant, and $e$ is the elementary charge. The total qubit capacitance is designed to be $C_\Sigma \approx 90$\,fF, yielding a charging energy $E_C /\hbar = e^2/(2C_\Sigma \times \hbar) \approx 2\pi \times 215$\,MHz. The resulting transmon frequency $\omega_{01} \approx (\sqrt{8E_J E_C} - E_C)/\hbar \approx 2\pi \times 4.85$\,GHz, with anharmonicity $\alpha = \omega_{12} - \omega_{01} \approx -E_C /\hbar$ \cite{Koch2007-vs}. This ensures the qubit operates in the transmon regime ($E_J/E_C > 60$) with suppressed sensitivity to charge noise.

\begin{figure*}[!ht]
    \centering
    \includegraphics[width=0.8\linewidth]{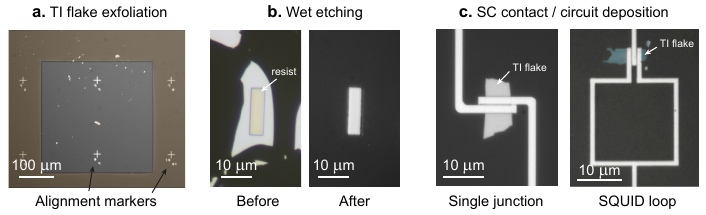}
    \caption{Fabrication. (a) BSTS flakes are exfoliated onto the sapphire substrate pre-patterned with alignment markers. We cover the substrate with protective resist, except for a central square region, to facilitate the removal of unwanted flakes outside this target region. The resist is indicated in yellow false color. (b) Shaping of flakes. The desired shape is defined via lithography, followed by an acid etch. (c) Deposition of SC contacts and circuits. \textit{Left:} Single S-TI-S planar junction formed by MoRe contacts on unshaped BSTS flake. \textit{Right:} SQUID loop by forming two S-TI-S planar junctions on a single TI flake, shown in blue false color.
    } 
    \label{fig:2a-fab}
\end{figure*}

The designed $I_c$ is achievable with a junction gap of $\sim 100$\,nm and a lateral junction length of $2–10\,\mu$m. The design tolerates significant fabrication variability: for $I_c$ in the range of $10–60$\,nA and keeping all other parameters constant, the qubit frequency varies from $2\pi \times 2.7$\,GHz to $2\pi \times 6.9$\,GHz, which can be measured with the same microwave cavity and hardware setup. Alternatively, a superconducting interference device (SQUID) can be fabricated using two S-TI-S junctions to enable tunability of $I_c$ and $\omega_{01}$ (see later discussion on fabrication and Fig.\,\ref{fig:2a-fab}).
The total qubit capacitance arises from three components: the inter-island capacitance $C_{\text{L–R}}$, the capacitance from each island to the cavity walls $C_{\text{L–C}}$, and the junction capacitance $C_J$. From the effective circuit shown in Fig.\,\ref{fig:1-design}b, the effective total capacitance is $C_\Sigma = C_\text{L-R} + 1/(1/C_\text{L-C}+1/C_\text{L-C}) + C_J$, which is shown as the circuit diagram in the dashed-line box. 

\begin{figure*}[!ht]
    \centering
    \includegraphics[width=1.0\linewidth]{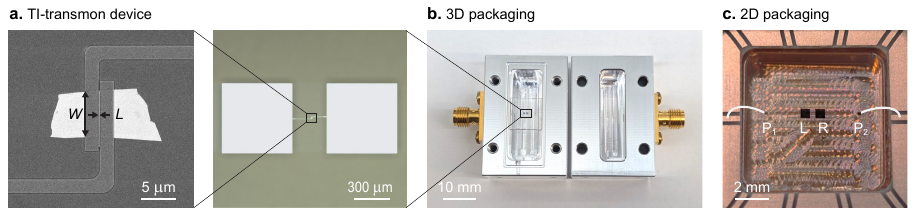}
    \caption{Device and packaging. (a) Optical image of the TI-transmon device showing the split capacitor pads and SEM image of the S-TI-S junction. (b) Device chip mounted in the aluminum 3D cavity. (c) Device mounted to a custom PCB and copper enclosure, as illustrated here with wire bonds connected to the small coupling pads for RF spectroscopy of the TI-transmon.
    }
    \label{fig:2b-package}
\end{figure*}

The transmon is fabricated on a 10$\times$10\,mm sapphire substrate and mounted inside a 3D aluminum cavity with dimensions 28$\times$27$\times$8\,mm (see later discussion on packaging and Fig.\,\ref{fig:2b-package}). In the absence of any device inside, the cavity's fundamental mode ($\text{TE}_{110}$) has a frequency $\omega_r \approx 2\pi \times 7.7$\,GHz \cite{Pozar2012-ad}. In the presence of the sapphire substrate, the dielectric environment inside the cavity is modified, reducing the cavity mode frequency to approximately $2\pi \times 7.1$\,GHz.
Using a finite element electromagnetic solver (Ansys Q3D), we extract the geometry-dependent capacitances based on a design with two 600$\times$600\,$\mu$m SC pads separated by 300\,$\mu$m. The resulting capacitance values are: $C_{\text{L–R}} = 39$\,fF, $C_{\text{L–C}} \approx C_{\text{R–C}} = 89$\,fF, and $C_J = 4$\,fF estimated from junction dimensions, giving a total $C_\Sigma \approx 90$\,fF.

The dipole moment of the TI-transmon capacitively couples to the electric field of the cavity mode.
This coupling enables standard circuit QED dispersive readout of the qubit when the qubit is far detuned in frequency from the cavity. The design above yields a coupling rate $g \approx 2\pi \times 215$\,MHz, which we verify in Ansys HFSS. The coupling can also be estimated from the effective electric dipole moment of the qubit and the electric field distribution of the cavity mode.
Importantly, the 3D architecture allows independent adjustment of the qubit capacitance $C_\Sigma$ and the qubit–cavity coupling $g$. The electric field of the qubit mode resides predominantly within the vacuum region, reducing participation in the dielectric substrate and decreasing decoherence from material losses.

Two coaxial RF SMA connectors, positioned on the cavity sidewalls, couple to the cavity field via antenna pins, enabling reflection or transmission measurements. The cavity-qubit electromagnetic response is modeled using Ansys HFSS to extract relavant circuit QED parameters. For additional design strategies for SC qubits, see reviews \cite{Gao2021-kj, Krantz2019-zv}.

\subsection{Unified design for characterization}

The development of hybrid SC qubits that incorporate novel materials often requires iterative optimization of design, materials, and fabrication processes. Clean, low-loss interfaces are essential for coherence, yet directly correlating material properties with qubit performance is often a challenging task.
To support such material–device co-design, we implement a unified architecture that allows for multiple types of measurements on the same physical device, including DC transport, RF spectroscopy, and circuit QED control and readout.

\textit{1. DC transport measurement.}
To characterize the S–TI–S junction independently, we perform two-wire DC transport measurements by wire bonding directly from the SMA connector center pins to the large capacitor pads of the transmon, illustrated in Fig.\,\ref{fig:1-design}c. This configuration enables measurement of the I-V characteristics of the junction at low temperatures. Following DC characterization, the wirebonds can be removed with minimal disturbance to the device, allowing subsequent RF measurements to proceed on the same sample.

\textit{2. Direct RF spectroscopy.} 
For direct spectroscopic characterization of the qubit, we include two additional small superconducting islands near the qubit capacitor pads, as shown in Fig.\,\ref{fig:1-design}d. These small coupling pads enable RF transmission measurements through input and output wirebonds, with minimal effect on the transmon parameters. Multiple sets of coupling pads with varying spacings to the qubit pad can be included to adjust $\kappa$ to match the intrinsic linewidth of the TI-transmon, thereby optimizing the signal-to-noise ratio for spectroscopy. We position the coupling pads approximately 1.5\,mm edge-to-edge from the qubit capacitor pads, with sizes of $200\times 200\,\mu$m, yielding a coupling capacitance of $\sim$ 3\,fF between the input/output pads and the TI-transmon. This translates to a coupling-induced qubit linewidth $\kappa_c \approx 2\pi \times 0.8$\,MHz. If the TI-transmon has an internal linewidth $\ll \kappa_c$, a Lorentzian transmission peak near $\omega_q$ is expected with peak transmission approaching 0\,dB. The spectroscopy allows us to directly probe the bare qubit frequency and verify other design parameters.

\textit{3. Circuit QED control and readout.}
Finally, after removing wirebonds from previous measurements, the TI-transmon chip is placed inside the 3D cavity for dispersive control and readout using standard circuit QED techniques (Fig.\,\ref{fig:1-design}a-b). The coupling between the cavity and external SMA connectors is determined by the length of the center pins. We employ asymmetric coupling with $\kappa_{\text{in}} \approx \kappa_{\text{out}}/5$ to enhance the readout signal-to-noise ratio. The total external coupling rate $\kappa_c = \kappa_{\text{in}} + \kappa_{\text{out}}$ is chosen to balance fast readout and suppression of Purcell decay \cite{Krantz2019-zv}. Since the internal quality factor of the SC cavity is much smaller compared to the desired $\kappa_c$, the total cavity linewidth $\kappa \approx \kappa_c$. To read out the qubit state within its lifetime $T_1$, the cavity must satisfy $1/\kappa \lesssim T_1$, while ensuring that the Purcell-limited lifetime $T_1^{\text{P}} \approx \Delta^2/\kappa g^2$ remains much longer than the intrinsic $T_1$. For our design parameters $g \approx 2\pi \times 185$\,MHz, detuning $\Delta = \omega_r - \omega_q \approx 2\pi \times 2$\,GHz, and $\kappa \approx 2\pi \times 0.5$\,MHz, we estimate $T_1^{\text{P}} \sim 40\,\mu$s, suitable for initial measurements of devices with $\mu$s-level coherence times. The length of the coupling pins and hence the input and output couplings can be easily adjusted to tailor to devices with higher intrinsic coherence times. For the circuit QED experiments, the qubit is driven near $\omega_q$ via the same SMA input port used for driving the readout cavity.

\section{Device fabrication\label{sec:fab}}

\textit{1. Substrate preparation.} We begin with double-side polished $c$-plane sapphire wafers (430\,$\mu$m thick, CrysTec GmbH), which are annealed in air at 1200$\degree$C for 2 hours to improve surface smoothness \cite{Pham-Van1998-ey, Kamal2016-bf}. The wafers are diced into $10 \times 10$\,mm chips and cleaned by sequential sonication in toluene, acetone, methanol, and isopropanol.

To facilitate subsequent fabrication steps, we pattern a grid of alignment markers using optical lithography. The markers are 10\,nm Al / 40\,nm Au bilayer (Al covered by Au) deposited via e-beam evaporation, followed by liftoff and resist removal in an 85$\degree$C N-methyl-2-pyrrolidone (NMP) bath. The Al layer promotes adhesion, while the Au layer enhances imaging contrast during e-beam lithography. As shown in Fig.\,\ref{fig:2a-fab}a, the markers are cross-shaped with unique identifiers, spaced at 200\,$\mu$m intervals optimized for our specific e-beam alignment.

\textit{2. TI exfoliation.}
High-quality single crystals of BSTS are grown via the Bridgman method \cite{Xu2014-fv}. BSTS is a layered 3D TI with a van der Waals structure, allowing mechanical exfoliation into thin flakes. The flakes are exfoliated using the standard scotch-tape technique and identified under an optical microscope. Candidate flakes are further characterized by atomic force microscopy to confirm thickness and surface flatness, with a typical desired thickness of $10-50$\,nm. The lateral sizes of the flakes range from a few microns to tens of microns.

For devices presented in this paper, BSTS flakes are exfoliated directly onto the sapphire substrate. To minimize unwanted flake deposition, an optimal lithography step may be used to open a small window on the chip before exfoliation (see Fig.\,\ref{fig:2a-fab}a). 
Alternatively, selected flakes can be transferred using a polymer-assisted technique \cite{Son2020-kw}, which allows for controlled placement and pre-selection based on thickness, size, or surface characteristics. The transfer method could minimize unwanted flake deposition on the device substrate, reducing the need for subsequent etching.

\textit{3. TI patterning and etching.}
To define the desired flake geometry to match the designed junction dimensions, we perform wet etching using a PMMA mask patterned by e-beam lithography, shown in Fig.\,\ref{fig:2a-fab}b. The resist is developed in an IPA:H\textsubscript{2}O (3:1) solution at 6$\degree$C for 3 minutes. The TI etching is carried out in a K\textsubscript{2}Cr\textsubscript{2}O\textsubscript{7}:H\textsubscript{2}SO\textsubscript{4}:H\textsubscript{2}O mixture (0.5\,g\,:\,2\,mL\,:\,10\,mL), with an etch rate of $\sim 20$\,nm/min for BSTS.  
After the TI flake etching, most of the alignment markers are removed to avoid microwave loss, except those near the chip corners which are used in the subsequent e-beam lithography of the junction contacts and SC circuits. The flake is protected with resist during this step. The markers' Au layer is removed using GE-8110 etchant, followed by Al removal in AZ 300 MIF developer. Residual resist is stripped with acetone, IPA, and DI water.

\textit{4. SC contacts and circuits.}
Our 3D TI-transmon design simplifies the fabrication of the remaining SC circuit, thereby providing a versatile platform to explore different SC contact materials. Here, we fabricate all SC elements from Molybdenum-Rhenium (MoRe), a material compatible with high-coherence SC circuits \cite{Singh2014-aw}.
The full circuit, including the junction leads and capacitor pads, is defined in a single e-beam lithography step. The minimum feature is the $\sim 100$\,nm junction gap, and the largest features are the qubit capacitor pads. We use single-layer PMMA with a conductive top resist layer (ESpacer 300Z) for e-beam exposure. The e-beam lithography is performed in an e-beam writer (JEOL 8100), followed by resist development as described above.
Prior to metal deposition, a gentle Ar ion milling step is used to remove resist residue and native oxide from the TI flake, etching away $1-2$\,nm from the top of the flake. MoRe is then sputtered at a rate of 0.4\,nm/s in an AJA deposition system at a base pressure of $2\times10^{-9}$\,mbar, followed by liftoff in NMP. An SEM image of a typical junction is shown in Fig.\,\ref{fig:2b-package}a, with junction width $W \approx 6.6\,\mu$m and length $L \approx 150$\,nm. Here, the BSTS flake has a thickness of 40\,nm.

\textit{5. Packaging.}
Figure \ref{fig:2b-package}a-b shows the completed TI-transmon chip mounted inside a 3D aluminum cavity. The chip is held in place by a thin indium layer along the edge, ensuring thermal contact and mechanical stability.
Alternatively, we can use a 2D PCB-based packaging for DC and RF spectroscopy measurements as shown in Fig.\,\ref{fig:2b-package}c. The chip is mounted and wirebonded to a custom sample PCB board and enclosed in a metal box with a fundamental mode above $\sim 12$\,GHz to suppress spurious coupling, similar to designs in \cite{Huang2021-oz}. The PCB-based package facilitates easier 4-wire DC measurements. For RF spectroscopy, the modified microwave environment from the enclosure alters the qubit’s charging energy $E_c$, leading to predictable shifts in qubit frequency that can be modeled and used to correlate design parameters with the measured spectra.

\section{Measurements\label{sec:results}}

\subsection{Critical current of S-TI-S junction}

\begin{figure*}[!ht]
    \centering
    \includegraphics[width=0.95\linewidth]{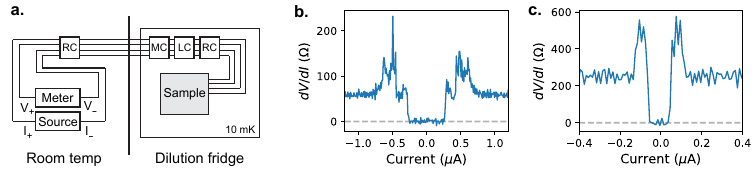}
    \caption{DC characterization of the S-TI-S junction. (a) Measurement setup. The DC signals are filtered by RC filters at room temperature, then by a Minicircuit low-pass filter (MC) and custom LC and RC filters at 10\,mK. (b,c) Differential resistance measurements of two S-BSTS-S junctions with different dimensions and hence different critical currents.}
    \label{fig:3-dc}
\end{figure*}

We begin by characterizing the DC transport properties of the S-TI-S junctions with 45\,nm MoRe superconducting contacts. The sputtered MoRe film exhibits a critical temperature $T_c \approx 8.8$\,K, measured independently.
We fabricated junctions with lateral widths $W = 5–30\,\mu$m and gap sizes $L = 80 – 200$\,nm. Figure~\ref{fig:3-dc}b-c shows DC measurement from two representative junctions: one with $W = 20\,\mu$m, $L = 110$\,nm, and TI thickness $T \approx 56$ \,nm; and another with $W = 9\,\mu$m, $L = 125$\,nm, $T \approx 49$ \,nm. All measurements are performed at zero magnetic field and without applying gate voltage.

The DC measurement setup is illustrated in Fig.~\ref{fig:3-dc}a. Samples are mounted on the mixing chamber plate of a dilution refrigerator at 10\,mK. DC signals are delivered via twisted-pair Cu wires to 4\,K, then via NbTi superconducting wires from 4\,K to the mixing chamber. At the mixing chamber plate, each line is filtered by a low-pass LC filter and a custom RC filter with infrared filtering. The RC filters feature a 340\,Hz cutoff using low-temperature NiCr resistors (1\,k$\Omega$) and thin-film capacitors (0.47\,$\mu$F). Each signal line is routed through a meandering trace on a custom copper PCB covered by an infrared absorber (Eccosorb CRS-117).
The room temperature DC transport measurements are performed using a precision DC source (Yokogawa GS200) and a low-noise digital multimeter (Keithley 2010), with additional RC filters on each line (with 100\,kHz cutoff). The differential resistance data were taken with a 100\,ms integration time and averaged over 100 repetitions for each current value. All observed $dV/dI$ traces are symmetric about zero bias and exhibit no hysteresis between forward and reverse current sweeps.

For the larger junction ($W = 25,\mu$m, Fig.~\ref{fig:3-dc}b), we measure a critical supercurrent $I_c \approx 250$\,nA. Above $I_c$, a transition region extends from 250\,nA to 650\,nA in which $dV/dI$ exhibits oscillatory features symmetric around $I = 0\,\mu$A, suggesting the presence of sub-gap transport processes \cite{Oostinga2013-xi}. The junction normal-state resistance is $R_n \approx 60\,\Omega$, yielding an $I_c R_n$ product of $\approx 15\,\mu$V. This measured $I_c R_n$ is significantly smaller than the BCS superconducting gap of MoRe ($\Delta_{\mathrm{MoRe}} \approx 1.76\,k_B T_c \approx 1.33$\,meV), indicating low effective junction transparency. This low transparency could arise from factors such as interface quality, disorder, and/or be characteristic of transport dominated by the topological surface states. Further studies, such as characterizing multiple Andreev reflections \cite{Jauregui2018-fr}, are needed to elucidate the origin. In the smaller junction ($W = 8\,\mu$m, Fig.~\ref{fig:3-dc}c), we measure $I_c \approx 45$\,nA and $R_n \approx 250\,\Omega$ ($I_c R_n \approx 11\,\mu$V), with the transition region extending to $\sim 150$\,nA. These measurements have guided our junction design and fabrication, allowing us to achieve the target $I_c$ for the TI-transmon by adjusting the junction width $W$.

\subsection{RF Measurement Setup}

Here we describe the RF measurement setup we will use in future work for full circuit QED control and readout of the hybrid TI-transmon qubit.
 
The circuit QED characterization of SC qubit devices typically begins with frequency-domain measurements: Single-tone spectroscopy of the resonator and two-tone spectroscopy of the qubit are performed using a vector network analyzer (VNA). These are followed by time-domain control and measurement experiments using custom or commercial setups comprising arbitrary waveform generators, digitizers, RF sources, and mixers \cite{Gao2021-kj}.

Recent advances in Radio Frequency System-on-Chip (RFSoC) technology have enabled quantum control systems with significantly lowered cost and complexity. RFSoCs provide real-time microwave pulse synthesis, high-speed analog-to-digital conversion, and low-latency signal processing via field-programmable gate arrays (FPGAs). Here, we use the open-source Quantum Instrumentation Control Kit (QICK) \cite{Stefanazzi2022-gx, Ding2024-vz} built on Xilinx RFSoC hardware. For single-qubit control, the Xilinx 4x2 educational evaluation board \cite{AMD-RFSoC4x2} offers a compact and accessible option. When paired with a programmable bandpass filter on the DAC output, the Xilinx RFSoC board can also serve as a cost-effective alternative to a VNA.

\begin{figure*}[!ht]
    \centering
    \includegraphics[width=0.67\linewidth]{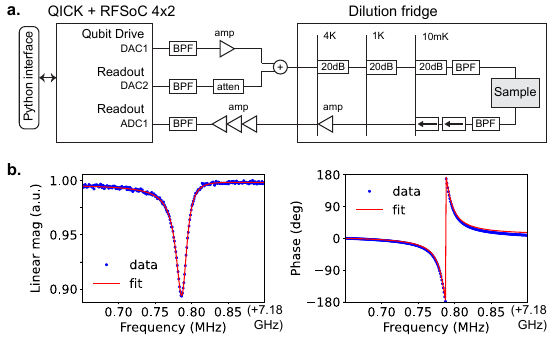}
    \caption{RF measurement. (a) Microwave and cryogenic wiring setup, using a Xilinx RFSoC 4x2 system running QICK. (b) Reflection spectrum of the aluminum cavity mounted with a blank sapphire chip, showing the measured linear magnitude and phase of the complex $S_{11}$ signal and the fitting result.}
    \label{fig:4-rf}
\end{figure*}

Figure~\ref{fig:4-rf} shows the full RF setup, including both room-temperature and cryogenic components. The input RF signal is attenuated by more than 60\,dB across multiple temperature stages before reaching the device at base temperature. The output signal passes through two cryogenic isolators (Quantum Microwave) and is amplified at the 4\,K stage by a high-electron-mobility transistor (HEMT, Low Noise Factory). Additional low-pass filters (e.g. Marki FLP0960) and infrared filters (Eccosorb) are used at the mixing chamber to suppress high-frequency noise from warmer stages of the fridge.
Running standard QICK firmware, two DAC channels of the RFSoC 4x2 with sampling rate 4.9\,GS/s will be used to synthesize the qubit drive and resonator readout pulses. These signals are band-pass-filtered (BPF in Fig.\,~\ref{fig:4-rf}), attenuated or amplified as needed, and combined onto a single drive line with a power splitter. The amplified output signal from the fridge is filtered and then digitized by one of the high-speed ADC channels on the RFSoC board with a sampling rate of $2.5$\,GS/s.

\subsection{Microwave loss characterization \\using high-Q superconducting cavity}

To investigate potential sources of microwave loss in our hybrid devices, we performed microwave characterization using the high-quality factor (high-Q) aluminum 3D cavities for different steps involved in the device fabrication process. We measure cavity reflection using a single weakly-coupled port to enhance sensitivity to changes in the internal quality factor $Q_i$ of the cavity mode due to the presence of the TI-transmon chip, here using a Keysight VNA. Open-source analysis tools are available to systematically extract the quality factors from cavity spectra \cite{McRae2020-ak, scresonators, Wang2021-lv}.

We begin by mounting a clean blank sapphire chip inside the aluminum cavity and measuring the resulting internal quality factor. 
The cavity reflection spectrum $S_{11}(\omega)$ is measured via a single port coupled to the cavity and using a directional coupler to route the reflection signal towards to output line in the fridge. The complex reflection voltage signal $S_{11}(\omega)$, shown in Fig.\,\ref{fig:4-rf}b, is fitted to the following functional form: 
\begin{equation*}
    S_{11}(\omega) = A\,e^{i\theta} \left[1- \frac{\kappa_c\,e^{i\phi}}{i(\omega-\omega_r) + (\kappa_c+\kappa_i)/2} \right].
\end{equation*}
Here $A\,e^{i\theta}$ is a complex background that accounts for attenuation and phase delay from the wiring, taken to be frequency-independent over the narrow frequency range around the cavity resonance. $\omega_r$ is the cavity resonance frequency. $\kappa_c$ and $\kappa_i$ are the cavity external (coupling) linewidth and internal (intrinsic) linewidth, corresponding to an external (coupling) quality factor $Q_c = \omega_r/\kappa_c$ and internal quality factor $Q_i = \omega_r/\kappa_i$. The angle $\phi$ captures the asymmetric resonance line shape arising from impedance mismatch or stray coupling in the input/ouput transmission lines \cite{Khalil2012-yi}, with $\phi=0$ corresponding to a symmetric Lorentzian shape.
From the fitting (Fig.\,\ref{fig:4-rf}b), we extract a resonance frequency $\omega_r \approx 2\pi\times 7.1808$\,GHz, external quality factor $Q_c \approx 4.1\times 10^5$ ($\kappa_c \approx 2\pi\times 17$\,kHz), and internal quality factor $Q_i \approx 7.7\times 10^6$ ($\kappa_i \approx 2\pi\times 0.9$\,kHz).
The cavity is therefore overcoupled ($\kappa_c>\kappa_i$), as can be seen from the $360^\circ$ phase difference across the resonance.
Next, we introduced ten exfoliated TI flakes onto the chip and repeated the same measurement to observe a reduced cavity internal quality factor of $Q_i \approx 9\times 10^5$. After removing the flakes via the wet etching process described above, $Q_i$ recovered to its original value ($\approx 8\times10^6$), suggesting that the etching procedure does not introduce significant additional loss.
To further isolate loss mechanisms, we can examine intermediate configurations. For instance, a sapphire chip with MoRe circuit structures but no TI flakes would allow us to assess loss contributions from the MoRe SC film. Assuming a uniform loss tangent across the MoRe layer, we can place an upper bound on the quality factor of the qubit mode by comparing the energy participation ratios of the cavity and qubit modes in the lossy SC film \cite{Minev2021-dr}.

Additional spatial information about loss sources can be obtained by measuring quality factors of different modes of the 3D cavity. In future work, we plan to implement planar superconducting resonators to characterize material and interface losses by terminating coplanar waveguide resonators with S-TI-S junctions \cite{Bottcher2024-kc}.

\section{Conclusion and outlook\label{sec:conclusion}}

We have presented the design and experimental progress toward a hybrid TI-transmon qubit incorporating $\mathrm{BiSbTeSe_2}$ topological insulator junctions in a 3D circuit QED architecture. Our approach provides a unified framework for performing DC transport, RF spectroscopy, and circuit-QED measurements on a single device, facilitating the efficient co-design of materials, fabrication processes, and qubit architectures for hybrid quantum devices.

As a natural next step, the integration of electrostatic gates \cite{Xia2024-dz} will enable in situ voltage control of the critical current of the S-TI-S junction, resulting in fast qubit frequency tuning \cite{Jauregui2018-fr}. The incorporation of on-chip LC low-pass filters on the gate control line can help mitigate Purcell decay and enhance coherence times.
Adding flux tunability via a SQUID loop is another extension. This could be realized using either two S-TI-S junctions or a combination of an S-TI-S junction with a conventional aluminum junction. Recent studies have demonstrated operation of 3D devices with aluminum junctions under moderate magnetic fields \cite{Krause2022-ap, Krause2024-xg}, opening the door to experiments that explore field-tunable effects in hybrid SC qubit devices with topological materials.

Looking ahead, hybrid superconducting qubits based on topological insulator junctions may offer noise resilience that stems from the nontrivial current-phase relation and parity-symmetry protection \cite{Larsen2020-fi, Guo2024-mo}, or from engineered dispersion in multi-junction arrays \cite{Doucot2012-ax, Gyenis2021-vd}.
More broadly, the hybrid platform offers a promising route for probing exotic topological excitations using superconducting circuits and circuit QED toolbox \cite{Hassler2011-zr, van-Heck2015-en}. Multi-terminal S-TI-S junctions have been proposed as a platform for detecting and manipulating Majorana bound states \cite{Hegde2020-nw}, with recent experiments beginning to uncover relevant signatures \cite{Yue2024-nx}. Advancement in these hybrid SC circuit devices could ultimately enable new functionalities for quantum information processing and novel opportunities to explore many-body physics in engineered topological systems.

\section*{Acknowledgment}

The authors thank Peter Salisbury for designing and constructing the custom low-pass filters used in this work.
This work was supported by the U.S. Department of Energy, Office of Science, National Quantum Information Science Research Centers, Quantum Science Center.

\vspace{0.25in}
\textbf{Data availability statement:} The data that support the findings of this study are available from the corresponding author upon reasonable request.

\bibliography{references}

\begin{thebibliography}{63}%
\makeatletter
\providecommand \@ifxundefined [1]{%
 \@ifx{#1\undefined}
}%
\providecommand \@ifnum [1]{%
 \ifnum #1\expandafter \@firstoftwo
 \else \expandafter \@secondoftwo
 \fi
}%
\providecommand \@ifx [1]{%
 \ifx #1\expandafter \@firstoftwo
 \else \expandafter \@secondoftwo
 \fi
}%
\providecommand \natexlab [1]{#1}%
\providecommand \enquote  [1]{``#1''}%
\providecommand \bibnamefont  [1]{#1}%
\providecommand \bibfnamefont [1]{#1}%
\providecommand \citenamefont [1]{#1}%
\providecommand \href@noop [0]{\@secondoftwo}%
\providecommand \href [0]{\begingroup \@sanitize@url \@href}%
\providecommand \@href[1]{\@@startlink{#1}\@@href}%
\providecommand \@@href[1]{\endgroup#1\@@endlink}%
\providecommand \@sanitize@url [0]{\catcode `\\12\catcode `\$12\catcode `\&12\catcode `\#12\catcode `\^12\catcode `\_12\catcode `\%12\relax}%
\providecommand \@@startlink[1]{}%
\providecommand \@@endlink[0]{}%
\providecommand \url  [0]{\begingroup\@sanitize@url \@url }%
\providecommand \@url [1]{\endgroup\@href {#1}{\urlprefix }}%
\providecommand \urlprefix  [0]{URL }%
\providecommand \Eprint [0]{\href }%
\providecommand \doibase [0]{https://doi.org/}%
\providecommand \selectlanguage [0]{\@gobble}%
\providecommand \bibinfo  [0]{\@secondoftwo}%
\providecommand \bibfield  [0]{\@secondoftwo}%
\providecommand \translation [1]{[#1]}%
\providecommand \BibitemOpen [0]{}%
\providecommand \bibitemStop [0]{}%
\providecommand \bibitemNoStop [0]{.\EOS\space}%
\providecommand \EOS [0]{\spacefactor3000\relax}%
\providecommand \BibitemShut  [1]{\csname bibitem#1\endcsname}%
\let\auto@bib@innerbib\@empty
\bibitem [{\citenamefont {Kjaergaard}\ \emph {et~al.}(2020)\citenamefont {Kjaergaard}, \citenamefont {Schwartz}, \citenamefont {Braumüller}, \citenamefont {Krantz}, \citenamefont {Wang}, \citenamefont {Gustavsson},\ and\ \citenamefont {Oliver}}]{Kjaergaard2020-rt}%
  \BibitemOpen
  \bibfield  {author} {\bibinfo {author} {\bibfnamefont {M.}~\bibnamefont {Kjaergaard}}, \bibinfo {author} {\bibfnamefont {M.~E.}\ \bibnamefont {Schwartz}}, \bibinfo {author} {\bibfnamefont {J.}~\bibnamefont {Braumüller}}, \bibinfo {author} {\bibfnamefont {P.}~\bibnamefont {Krantz}}, \bibinfo {author} {\bibfnamefont {J.~I.-J.}\ \bibnamefont {Wang}}, \bibinfo {author} {\bibfnamefont {S.}~\bibnamefont {Gustavsson}},\ and\ \bibinfo {author} {\bibfnamefont {W.~D.}\ \bibnamefont {Oliver}},\ }\bibfield  {title} {\bibinfo {title} {Superconducting qubits: Current state of play},\ }\href {https://doi.org/10.1146/annurev-conmatphys-} {\bibfield  {journal} {\bibinfo  {journal} {Annu. Rev. Condens. Matter Phys.}\ }\textbf {\bibinfo {volume} {11}},\ \bibinfo {pages} {369} (\bibinfo {year} {2020})}\BibitemShut {NoStop}%
\bibitem [{\citenamefont {Blais}\ \emph {et~al.}(2020)\citenamefont {Blais}, \citenamefont {Girvin},\ and\ \citenamefont {Oliver}}]{Blais2020-ef}%
  \BibitemOpen
  \bibfield  {author} {\bibinfo {author} {\bibfnamefont {A.}~\bibnamefont {Blais}}, \bibinfo {author} {\bibfnamefont {S.~M.}\ \bibnamefont {Girvin}},\ and\ \bibinfo {author} {\bibfnamefont {W.~D.}\ \bibnamefont {Oliver}},\ }\bibfield  {title} {\bibinfo {title} {Quantum information processing and quantum optics with circuit quantum electrodynamics},\ }\href {https://doi.org/10.1038/s41567-020-0806-z} {\bibfield  {journal} {\bibinfo  {journal} {Nat. Phys.}\ }\textbf {\bibinfo {volume} {16}},\ \bibinfo {pages} {247} (\bibinfo {year} {2020})}\BibitemShut {NoStop}%
\bibitem [{\citenamefont {Carusotto}\ \emph {et~al.}(2020)\citenamefont {Carusotto}, \citenamefont {Houck}, \citenamefont {Kollár}, \citenamefont {Roushan}, \citenamefont {Schuster},\ and\ \citenamefont {Simon}}]{Carusotto2020-ct}%
  \BibitemOpen
  \bibfield  {author} {\bibinfo {author} {\bibfnamefont {I.}~\bibnamefont {Carusotto}}, \bibinfo {author} {\bibfnamefont {A.~A.}\ \bibnamefont {Houck}}, \bibinfo {author} {\bibfnamefont {A.~J.}\ \bibnamefont {Kollár}}, \bibinfo {author} {\bibfnamefont {P.}~\bibnamefont {Roushan}}, \bibinfo {author} {\bibfnamefont {D.~I.}\ \bibnamefont {Schuster}},\ and\ \bibinfo {author} {\bibfnamefont {J.}~\bibnamefont {Simon}},\ }\bibfield  {title} {\bibinfo {title} {Photonic materials in circuit quantum electrodynamics},\ }\href {https://www.nature.com/articles/s41567-020-0815-y} {\bibfield  {journal} {\bibinfo  {journal} {Nat. Phys.}\ }\textbf {\bibinfo {volume} {16}},\ \bibinfo {pages} {268} (\bibinfo {year} {2020})}\BibitemShut {NoStop}%
\bibitem [{\citenamefont {Antony}\ \emph {et~al.}(2021)\citenamefont {Antony}, \citenamefont {Gustafsson}, \citenamefont {Ribeill}, \citenamefont {Ware}, \citenamefont {Rajendran}, \citenamefont {Govia}, \citenamefont {Ohki}, \citenamefont {Taniguchi}, \citenamefont {Watanabe}, \citenamefont {Hone},\ and\ \citenamefont {Fong}}]{Antony2021-tf}%
  \BibitemOpen
  \bibfield  {author} {\bibinfo {author} {\bibfnamefont {A.}~\bibnamefont {Antony}}, \bibinfo {author} {\bibfnamefont {M.~V.}\ \bibnamefont {Gustafsson}}, \bibinfo {author} {\bibfnamefont {G.~J.}\ \bibnamefont {Ribeill}}, \bibinfo {author} {\bibfnamefont {M.}~\bibnamefont {Ware}}, \bibinfo {author} {\bibfnamefont {A.}~\bibnamefont {Rajendran}}, \bibinfo {author} {\bibfnamefont {L.~C.~G.}\ \bibnamefont {Govia}}, \bibinfo {author} {\bibfnamefont {T.~A.}\ \bibnamefont {Ohki}}, \bibinfo {author} {\bibfnamefont {T.}~\bibnamefont {Taniguchi}}, \bibinfo {author} {\bibfnamefont {K.}~\bibnamefont {Watanabe}}, \bibinfo {author} {\bibfnamefont {J.}~\bibnamefont {Hone}},\ and\ \bibinfo {author} {\bibfnamefont {K.~C.}\ \bibnamefont {Fong}},\ }\bibfield  {title} {\bibinfo {title} {Miniaturizing transmon qubits using van der waals materials},\ }\href
  {https://pubs.acs.org/doi/abs/10.1021/acs.nanolett.1c04160?casa_token=gAeVKN0-c9QAAAAA:twXXpGZes7uvpCZbvRGXbSGkIg3OEbN6T6Ez9HsM_zYnF_aUezz8Nv7n-pOoD_8p1fhoCWAHjzkUC6fjuw&casa_token=BtXQuaHwBI8AAAAA:CK3a7nbwwp5_jLMSsu5ybNRza3qJc_5jzI-LVOzBkD-XIon8b6stQobcK-JB0j9TmuYiK_IqSy3yt24Bvw} {\bibfield  {journal} {\bibinfo  {journal} {Nano Lett.}\ }\textbf {\bibinfo {volume} {21}},\ \bibinfo {pages} {10122} (\bibinfo {year} {2021})}\BibitemShut {NoStop}%
\bibitem [{\citenamefont {Wang}\ \emph {et~al.}(2022)\citenamefont {Wang}, \citenamefont {Yamoah}, \citenamefont {Li}, \citenamefont {Karamlou}, \citenamefont {Dinh}, \citenamefont {Kannan}, \citenamefont {Braumüller}, \citenamefont {Kim}, \citenamefont {Melville}, \citenamefont {Muschinske}, \citenamefont {Niedzielski}, \citenamefont {Serniak}, \citenamefont {Sung}, \citenamefont {Winik}, \citenamefont {Yoder}, \citenamefont {Schwartz}, \citenamefont {Watanabe}, \citenamefont {Taniguchi}, \citenamefont {Orlando}, \citenamefont {Gustavsson}, \citenamefont {Jarillo-Herrero},\ and\ \citenamefont {Oliver}}]{Wang2022-qn}%
  \BibitemOpen
  \bibfield  {author} {\bibinfo {author} {\bibfnamefont {J.~I.-J.}\ \bibnamefont {Wang}}, \bibinfo {author} {\bibfnamefont {M.~A.}\ \bibnamefont {Yamoah}}, \bibinfo {author} {\bibfnamefont {Q.}~\bibnamefont {Li}}, \bibinfo {author} {\bibfnamefont {A.~H.}\ \bibnamefont {Karamlou}}, \bibinfo {author} {\bibfnamefont {T.}~\bibnamefont {Dinh}}, \bibinfo {author} {\bibfnamefont {B.}~\bibnamefont {Kannan}}, \bibinfo {author} {\bibfnamefont {J.}~\bibnamefont {Braumüller}}, \bibinfo {author} {\bibfnamefont {D.}~\bibnamefont {Kim}}, \bibinfo {author} {\bibfnamefont {A.~J.}\ \bibnamefont {Melville}}, \bibinfo {author} {\bibfnamefont {S.~E.}\ \bibnamefont {Muschinske}}, \bibinfo {author} {\bibfnamefont {B.~M.}\ \bibnamefont {Niedzielski}}, \bibinfo {author} {\bibfnamefont {K.}~\bibnamefont {Serniak}}, \bibinfo {author} {\bibfnamefont {Y.}~\bibnamefont {Sung}}, \bibinfo {author} {\bibfnamefont {R.}~\bibnamefont {Winik}}, \bibinfo {author} {\bibfnamefont {J.~L.}\ \bibnamefont {Yoder}}, \bibinfo {author} {\bibfnamefont
  {M.~E.}\ \bibnamefont {Schwartz}}, \bibinfo {author} {\bibfnamefont {K.}~\bibnamefont {Watanabe}}, \bibinfo {author} {\bibfnamefont {T.}~\bibnamefont {Taniguchi}}, \bibinfo {author} {\bibfnamefont {T.~P.}\ \bibnamefont {Orlando}}, \bibinfo {author} {\bibfnamefont {S.}~\bibnamefont {Gustavsson}}, \bibinfo {author} {\bibfnamefont {P.}~\bibnamefont {Jarillo-Herrero}},\ and\ \bibinfo {author} {\bibfnamefont {W.~D.}\ \bibnamefont {Oliver}},\ }\bibfield  {title} {\bibinfo {title} {Hexagonal boron nitride as a low-loss dielectric for superconducting quantum circuits and qubits},\ }\href {https://doi.org/10.1038/s41563-021-01187-w} {\bibfield  {journal} {\bibinfo  {journal} {Nat. Mater.}\ }\textbf {\bibinfo {volume} {21}},\ \bibinfo {pages} {398} (\bibinfo {year} {2022})}\BibitemShut {NoStop}%
\bibitem [{\citenamefont {Tian}\ \emph {et~al.}(2021)\citenamefont {Tian}, \citenamefont {Jauregui}, \citenamefont {Wilen}, \citenamefont {Rigosi}, \citenamefont {Newell}, \citenamefont {McDermott},\ and\ \citenamefont {Chen}}]{Tian2021-mu}%
  \BibitemOpen
  \bibfield  {author} {\bibinfo {author} {\bibfnamefont {J.}~\bibnamefont {Tian}}, \bibinfo {author} {\bibfnamefont {L.~A.}\ \bibnamefont {Jauregui}}, \bibinfo {author} {\bibfnamefont {C.~D.}\ \bibnamefont {Wilen}}, \bibinfo {author} {\bibfnamefont {A.~F.}\ \bibnamefont {Rigosi}}, \bibinfo {author} {\bibfnamefont {D.~B.}\ \bibnamefont {Newell}}, \bibinfo {author} {\bibfnamefont {R.}~\bibnamefont {McDermott}},\ and\ \bibinfo {author} {\bibfnamefont {Y.~P.}\ \bibnamefont {Chen}},\ }\bibfield  {title} {\bibinfo {title} {A josephson junction with h-{BN} tunnel barrier: observation of low critical current noise},\ }\href {https://iopscience.iop.org/article/10.1088/1361-648X/ac268f/meta?casa_token=m8YDp51ZT7oAAAAA:ME6YF6GKh9HOIQkFyDoopGDqKEePS4Kc0A3EmHlXBhmBnKV4kexXWvNUsljRA_-e0DaqLqyBZ8vEq1VfYXuysrzVRWihpw} {\bibfield  {journal} {\bibinfo  {journal} {J. Phys. Condens. Matter}\ }\textbf {\bibinfo {volume} {33}},\ \bibinfo {pages} {495301} (\bibinfo {year} {2021})}\BibitemShut {NoStop}%
\bibitem [{\citenamefont {Larsen}\ \emph {et~al.}(2015)\citenamefont {Larsen}, \citenamefont {Petersson}, \citenamefont {Kuemmeth}, \citenamefont {Jespersen}, \citenamefont {Krogstrup}, \citenamefont {Nygård},\ and\ \citenamefont {Marcus}}]{Larsen2015-hu}%
  \BibitemOpen
  \bibfield  {author} {\bibinfo {author} {\bibfnamefont {T.~W.}\ \bibnamefont {Larsen}}, \bibinfo {author} {\bibfnamefont {K.~D.}\ \bibnamefont {Petersson}}, \bibinfo {author} {\bibfnamefont {F.}~\bibnamefont {Kuemmeth}}, \bibinfo {author} {\bibfnamefont {T.~S.}\ \bibnamefont {Jespersen}}, \bibinfo {author} {\bibfnamefont {P.}~\bibnamefont {Krogstrup}}, \bibinfo {author} {\bibfnamefont {J.}~\bibnamefont {Nygård}},\ and\ \bibinfo {author} {\bibfnamefont {C.~M.}\ \bibnamefont {Marcus}},\ }\bibfield  {title} {\bibinfo {title} {Semiconductor-nanowire-based superconducting qubit},\ }\href {http://link.aps.org/pdf/10.1103/PhysRevLett.115.127001} {\bibfield  {journal} {\bibinfo  {journal} {Phys. Rev. Lett.}\ }\textbf {\bibinfo {volume} {115}},\ \bibinfo {pages} {127001} (\bibinfo {year} {2015})}\BibitemShut {NoStop}%
\bibitem [{\citenamefont {de~Lange}\ \emph {et~al.}(2015)\citenamefont {de~Lange}, \citenamefont {van Heck}, \citenamefont {Bruno}, \citenamefont {van Woerkom}, \citenamefont {Geresdi}, \citenamefont {Plissard}, \citenamefont {Bakkers}, \citenamefont {Akhmerov},\ and\ \citenamefont {DiCarlo}}]{de-Lange2015-de}%
  \BibitemOpen
  \bibfield  {author} {\bibinfo {author} {\bibfnamefont {G.}~\bibnamefont {de~Lange}}, \bibinfo {author} {\bibfnamefont {B.}~\bibnamefont {van Heck}}, \bibinfo {author} {\bibfnamefont {A.}~\bibnamefont {Bruno}}, \bibinfo {author} {\bibfnamefont {D.~J.}\ \bibnamefont {van Woerkom}}, \bibinfo {author} {\bibfnamefont {A.}~\bibnamefont {Geresdi}}, \bibinfo {author} {\bibfnamefont {S.~R.}\ \bibnamefont {Plissard}}, \bibinfo {author} {\bibfnamefont {E.~P. A.~M.}\ \bibnamefont {Bakkers}}, \bibinfo {author} {\bibfnamefont {A.~R.}\ \bibnamefont {Akhmerov}},\ and\ \bibinfo {author} {\bibfnamefont {L.}~\bibnamefont {DiCarlo}},\ }\bibfield  {title} {\bibinfo {title} {Realization of microwave quantum circuits using hybrid superconducting-semiconducting nanowire josephson elements},\ }\href {http://link.aps.org/pdf/10.1103/PhysRevLett.115.127002} {\bibfield  {journal} {\bibinfo  {journal} {Phys. Rev. Lett.}\ }\textbf {\bibinfo {volume} {115}},\ \bibinfo {pages} {127002} (\bibinfo {year} {2015})}\BibitemShut {NoStop}%
\bibitem [{\citenamefont {Casparis}\ \emph {et~al.}(2018)\citenamefont {Casparis}, \citenamefont {Connolly}, \citenamefont {Kjaergaard}, \citenamefont {Pearson}, \citenamefont {Kringhøj}, \citenamefont {Larsen}, \citenamefont {Kuemmeth}, \citenamefont {Wang}, \citenamefont {Thomas}, \citenamefont {Gronin}, \citenamefont {Gardner}, \citenamefont {Manfra}, \citenamefont {Marcus},\ and\ \citenamefont {Petersson}}]{Casparis2018-ar}%
  \BibitemOpen
  \bibfield  {author} {\bibinfo {author} {\bibfnamefont {L.}~\bibnamefont {Casparis}}, \bibinfo {author} {\bibfnamefont {M.~R.}\ \bibnamefont {Connolly}}, \bibinfo {author} {\bibfnamefont {M.}~\bibnamefont {Kjaergaard}}, \bibinfo {author} {\bibfnamefont {N.~J.}\ \bibnamefont {Pearson}}, \bibinfo {author} {\bibfnamefont {A.}~\bibnamefont {Kringhøj}}, \bibinfo {author} {\bibfnamefont {T.~W.}\ \bibnamefont {Larsen}}, \bibinfo {author} {\bibfnamefont {F.}~\bibnamefont {Kuemmeth}}, \bibinfo {author} {\bibfnamefont {T.}~\bibnamefont {Wang}}, \bibinfo {author} {\bibfnamefont {C.}~\bibnamefont {Thomas}}, \bibinfo {author} {\bibfnamefont {S.}~\bibnamefont {Gronin}}, \bibinfo {author} {\bibfnamefont {G.~C.}\ \bibnamefont {Gardner}}, \bibinfo {author} {\bibfnamefont {M.~J.}\ \bibnamefont {Manfra}}, \bibinfo {author} {\bibfnamefont {C.~M.}\ \bibnamefont {Marcus}},\ and\ \bibinfo {author} {\bibfnamefont {K.~D.}\ \bibnamefont {Petersson}},\ }\bibfield  {title} {\bibinfo {title} {Superconducting gatemon qubit based on a
  proximitized two-dimensional electron gas},\ }\href {https://www.nature.com/articles/s41565-018-0207-y} {\bibfield  {journal} {\bibinfo  {journal} {Nat. Nanotechnol.}\ }\textbf {\bibinfo {volume} {13}},\ \bibinfo {pages} {915} (\bibinfo {year} {2018})}\BibitemShut {NoStop}%
\bibitem [{\citenamefont {Kroll}\ \emph {et~al.}(2018)\citenamefont {Kroll}, \citenamefont {Uilhoorn}, \citenamefont {van~der Enden}, \citenamefont {de~Jong}, \citenamefont {Watanabe}, \citenamefont {Taniguchi}, \citenamefont {Goswami}, \citenamefont {Cassidy},\ and\ \citenamefont {Kouwenhoven}}]{Kroll2018-ix}%
  \BibitemOpen
  \bibfield  {author} {\bibinfo {author} {\bibfnamefont {J.~G.}\ \bibnamefont {Kroll}}, \bibinfo {author} {\bibfnamefont {W.}~\bibnamefont {Uilhoorn}}, \bibinfo {author} {\bibfnamefont {K.~L.}\ \bibnamefont {van~der Enden}}, \bibinfo {author} {\bibfnamefont {D.}~\bibnamefont {de~Jong}}, \bibinfo {author} {\bibfnamefont {K.}~\bibnamefont {Watanabe}}, \bibinfo {author} {\bibfnamefont {T.}~\bibnamefont {Taniguchi}}, \bibinfo {author} {\bibfnamefont {S.}~\bibnamefont {Goswami}}, \bibinfo {author} {\bibfnamefont {M.~C.}\ \bibnamefont {Cassidy}},\ and\ \bibinfo {author} {\bibfnamefont {L.~P.}\ \bibnamefont {Kouwenhoven}},\ }\bibfield  {title} {\bibinfo {title} {Magnetic field compatible circuit quantum electrodynamics with graphene josephson junctions},\ }\href {https://www.nature.com/articles/s41467-018-07124-x} {\bibfield  {journal} {\bibinfo  {journal} {Nat. Commun.}\ }\textbf {\bibinfo {volume} {9}},\ \bibinfo {pages} {4615} (\bibinfo {year} {2018})}\BibitemShut {NoStop}%
\bibitem [{\citenamefont {Wang}\ \emph {et~al.}(2019)\citenamefont {Wang}, \citenamefont {Rodan-Legrain}, \citenamefont {Bretheau}, \citenamefont {Campbell}, \citenamefont {Kannan}, \citenamefont {Kim}, \citenamefont {Kjaergaard}, \citenamefont {Krantz}, \citenamefont {Samach}, \citenamefont {Yan}, \citenamefont {Yoder}, \citenamefont {Watanabe}, \citenamefont {Taniguchi}, \citenamefont {Orlando}, \citenamefont {Gustavsson}, \citenamefont {Jarillo-Herrero},\ and\ \citenamefont {Oliver}}]{Wang2019-il}%
  \BibitemOpen
  \bibfield  {author} {\bibinfo {author} {\bibfnamefont {J.~I.-J.}\ \bibnamefont {Wang}}, \bibinfo {author} {\bibfnamefont {D.}~\bibnamefont {Rodan-Legrain}}, \bibinfo {author} {\bibfnamefont {L.}~\bibnamefont {Bretheau}}, \bibinfo {author} {\bibfnamefont {D.~L.}\ \bibnamefont {Campbell}}, \bibinfo {author} {\bibfnamefont {B.}~\bibnamefont {Kannan}}, \bibinfo {author} {\bibfnamefont {D.}~\bibnamefont {Kim}}, \bibinfo {author} {\bibfnamefont {M.}~\bibnamefont {Kjaergaard}}, \bibinfo {author} {\bibfnamefont {P.}~\bibnamefont {Krantz}}, \bibinfo {author} {\bibfnamefont {G.~O.}\ \bibnamefont {Samach}}, \bibinfo {author} {\bibfnamefont {F.}~\bibnamefont {Yan}}, \bibinfo {author} {\bibfnamefont {J.~L.}\ \bibnamefont {Yoder}}, \bibinfo {author} {\bibfnamefont {K.}~\bibnamefont {Watanabe}}, \bibinfo {author} {\bibfnamefont {T.}~\bibnamefont {Taniguchi}}, \bibinfo {author} {\bibfnamefont {T.~P.}\ \bibnamefont {Orlando}}, \bibinfo {author} {\bibfnamefont {S.}~\bibnamefont {Gustavsson}}, \bibinfo {author} {\bibfnamefont
  {P.}~\bibnamefont {Jarillo-Herrero}},\ and\ \bibinfo {author} {\bibfnamefont {W.~D.}\ \bibnamefont {Oliver}},\ }\bibfield  {title} {\bibinfo {title} {Coherent control of a hybrid superconducting circuit made with graphene-based van der waals heterostructures},\ }\href {https://doi.org/10.1038/s41565-018-0329-2} {\bibfield  {journal} {\bibinfo  {journal} {Nat. Nanotechnol.}\ }\textbf {\bibinfo {volume} {14}},\ \bibinfo {pages} {120} (\bibinfo {year} {2019})}\BibitemShut {NoStop}%
\bibitem [{\citenamefont {Chiu}\ \emph {et~al.}(2025)\citenamefont {Chiu}, \citenamefont {Chang}, \citenamefont {Lasrado}, \citenamefont {Lo}, \citenamefont {Chen}, \citenamefont {Hsu}, \citenamefont {Chen}, \citenamefont {Tsai}, \citenamefont {{Samina}}, \citenamefont {Lin},\ and\ \citenamefont {Ke}}]{Chiu2025-or}%
  \BibitemOpen
  \bibfield  {author} {\bibinfo {author} {\bibfnamefont {K.-L.}\ \bibnamefont {Chiu}}, \bibinfo {author} {\bibfnamefont {Y.}~\bibnamefont {Chang}}, \bibinfo {author} {\bibfnamefont {A.~J.}\ \bibnamefont {Lasrado}}, \bibinfo {author} {\bibfnamefont {C.-H.}\ \bibnamefont {Lo}}, \bibinfo {author} {\bibfnamefont {Y.-H.}\ \bibnamefont {Chen}}, \bibinfo {author} {\bibfnamefont {T.-Y.}\ \bibnamefont {Hsu}}, \bibinfo {author} {\bibfnamefont {Y.-C.}\ \bibnamefont {Chen}}, \bibinfo {author} {\bibfnamefont {Y.-C.}\ \bibnamefont {Tsai}}, \bibinfo {author} {\bibnamefont {{Samina}}}, \bibinfo {author} {\bibfnamefont {Y.-H.}\ \bibnamefont {Lin}},\ and\ \bibinfo {author} {\bibfnamefont {C.-T.}\ \bibnamefont {Ke}},\ }\bibfield  {title} {\bibinfo {title} {Integration of graphene-based superconducting quantum circuits in a three-dimensional cavity},\ }\href {http://dx.doi.org/10.1103/PhysRevApplied.23.034059} {\bibfield  {journal} {\bibinfo  {journal} {Phys. Rev. Appl.}\ }\textbf {\bibinfo {volume} {23}},\ \bibinfo {pages}
  {034059} (\bibinfo {year} {2025})}\BibitemShut {NoStop}%
\bibitem [{\citenamefont {Clerk}\ \emph {et~al.}(2020)\citenamefont {Clerk}, \citenamefont {Lehnert}, \citenamefont {Bertet}, \citenamefont {Petta},\ and\ \citenamefont {Nakamura}}]{Clerk2020-io}%
  \BibitemOpen
  \bibfield  {author} {\bibinfo {author} {\bibfnamefont {A.~A.}\ \bibnamefont {Clerk}}, \bibinfo {author} {\bibfnamefont {K.~W.}\ \bibnamefont {Lehnert}}, \bibinfo {author} {\bibfnamefont {P.}~\bibnamefont {Bertet}}, \bibinfo {author} {\bibfnamefont {J.~R.}\ \bibnamefont {Petta}},\ and\ \bibinfo {author} {\bibfnamefont {Y.}~\bibnamefont {Nakamura}},\ }\bibfield  {title} {\bibinfo {title} {Hybrid quantum systems with circuit quantum electrodynamics},\ }\href {https://www.nature.com/articles/s41567-020-0797-9} {\bibfield  {journal} {\bibinfo  {journal} {Nat. Phys.}\ }\textbf {\bibinfo {volume} {16}},\ \bibinfo {pages} {257} (\bibinfo {year} {2020})}\BibitemShut {NoStop}%
\bibitem [{\citenamefont {Blais}\ \emph {et~al.}(2021)\citenamefont {Blais}, \citenamefont {Grimsmo}, \citenamefont {Girvin},\ and\ \citenamefont {Wallraff}}]{Blais2021-bn}%
  \BibitemOpen
  \bibfield  {author} {\bibinfo {author} {\bibfnamefont {A.}~\bibnamefont {Blais}}, \bibinfo {author} {\bibfnamefont {A.~L.}\ \bibnamefont {Grimsmo}}, \bibinfo {author} {\bibfnamefont {S.~M.}\ \bibnamefont {Girvin}},\ and\ \bibinfo {author} {\bibfnamefont {A.}~\bibnamefont {Wallraff}},\ }\bibfield  {title} {\bibinfo {title} {Circuit quantum electrodynamics},\ }\href {https://journals.aps.org/rmp/abstract/10.1103/RevModPhys.93.025005} {\bibfield  {journal} {\bibinfo  {journal} {Rev. Mod. Phys.}\ }\textbf {\bibinfo {volume} {93}} (\bibinfo {year} {2021})}\BibitemShut {NoStop}%
\bibitem [{\citenamefont {Read}\ \emph {et~al.}(2023)\citenamefont {Read}, \citenamefont {Chapman}, \citenamefont {Lei}, \citenamefont {Curtis}, \citenamefont {Ganjam}, \citenamefont {Krayzman}, \citenamefont {Frunzio},\ and\ \citenamefont {Schoelkopf}}]{Read2023-qy}%
  \BibitemOpen
  \bibfield  {author} {\bibinfo {author} {\bibfnamefont {A.~P.}\ \bibnamefont {Read}}, \bibinfo {author} {\bibfnamefont {B.~J.}\ \bibnamefont {Chapman}}, \bibinfo {author} {\bibfnamefont {C.~U.}\ \bibnamefont {Lei}}, \bibinfo {author} {\bibfnamefont {J.~C.}\ \bibnamefont {Curtis}}, \bibinfo {author} {\bibfnamefont {S.}~\bibnamefont {Ganjam}}, \bibinfo {author} {\bibfnamefont {L.}~\bibnamefont {Krayzman}}, \bibinfo {author} {\bibfnamefont {L.}~\bibnamefont {Frunzio}},\ and\ \bibinfo {author} {\bibfnamefont {R.~J.}\ \bibnamefont {Schoelkopf}},\ }\bibfield  {title} {\bibinfo {title} {Precision measurement of the microwave dielectric loss of sapphire in the quantum regime with parts-per-billion sensitivity},\ }\href {https://journals.aps.org/prapplied/abstract/10.1103/PhysRevApplied.19.034064} {\bibfield  {journal} {\bibinfo  {journal} {Phys. Rev. Appl.}\ }\textbf {\bibinfo {volume} {19}} (\bibinfo {year} {2023})}\BibitemShut {NoStop}%
\bibitem [{\citenamefont {Checchin}\ \emph {et~al.}(2022)\citenamefont {Checchin}, \citenamefont {Frolov}, \citenamefont {Lunin}, \citenamefont {Grassellino},\ and\ \citenamefont {Romanenko}}]{Checchin2022-ek}%
  \BibitemOpen
  \bibfield  {author} {\bibinfo {author} {\bibfnamefont {M.}~\bibnamefont {Checchin}}, \bibinfo {author} {\bibfnamefont {D.}~\bibnamefont {Frolov}}, \bibinfo {author} {\bibfnamefont {A.}~\bibnamefont {Lunin}}, \bibinfo {author} {\bibfnamefont {A.}~\bibnamefont {Grassellino}},\ and\ \bibinfo {author} {\bibfnamefont {A.}~\bibnamefont {Romanenko}},\ }\bibfield  {title} {\bibinfo {title} {Measurement of the low-temperature loss tangent of high-resistivity silicon using a high- {Q} superconducting resonator},\ }\href {https://journals.aps.org/prapplied/abstract/10.1103/PhysRevApplied.18.034013} {\bibfield  {journal} {\bibinfo  {journal} {Phys. Rev. Appl.}\ }\textbf {\bibinfo {volume} {18}} (\bibinfo {year} {2022})}\BibitemShut {NoStop}%
\bibitem [{\citenamefont {Tabuchi}\ \emph {et~al.}(2014)\citenamefont {Tabuchi}, \citenamefont {Ishino}, \citenamefont {Ishikawa}, \citenamefont {Yamazaki}, \citenamefont {Usami},\ and\ \citenamefont {Nakamura}}]{Tabuchi2014-ij}%
  \BibitemOpen
  \bibfield  {author} {\bibinfo {author} {\bibfnamefont {Y.}~\bibnamefont {Tabuchi}}, \bibinfo {author} {\bibfnamefont {S.}~\bibnamefont {Ishino}}, \bibinfo {author} {\bibfnamefont {T.}~\bibnamefont {Ishikawa}}, \bibinfo {author} {\bibfnamefont {R.}~\bibnamefont {Yamazaki}}, \bibinfo {author} {\bibfnamefont {K.}~\bibnamefont {Usami}},\ and\ \bibinfo {author} {\bibfnamefont {Y.}~\bibnamefont {Nakamura}},\ }\bibfield  {title} {\bibinfo {title} {Hybridizing ferromagnetic magnons and microwave photons in the quantum limit},\ }\href {https://link.aps.org/doi/10.1103/PhysRevLett.113.083603} {\bibfield  {journal} {\bibinfo  {journal} {Phys. Rev. Lett.}\ }\textbf {\bibinfo {volume} {113}},\ \bibinfo {pages} {083603} (\bibinfo {year} {2014})}\BibitemShut {NoStop}%
\bibitem [{\citenamefont {Phan}\ \emph {et~al.}(2022)\citenamefont {Phan}, \citenamefont {Senior}, \citenamefont {Ghazaryan}, \citenamefont {Hatefipour}, \citenamefont {Strickland}, \citenamefont {Shabani}, \citenamefont {Serbyn},\ and\ \citenamefont {Higginbotham}}]{Phan2022-py}%
  \BibitemOpen
  \bibfield  {author} {\bibinfo {author} {\bibfnamefont {D.}~\bibnamefont {Phan}}, \bibinfo {author} {\bibfnamefont {J.}~\bibnamefont {Senior}}, \bibinfo {author} {\bibfnamefont {A.}~\bibnamefont {Ghazaryan}}, \bibinfo {author} {\bibfnamefont {M.}~\bibnamefont {Hatefipour}}, \bibinfo {author} {\bibfnamefont {W.~M.}\ \bibnamefont {Strickland}}, \bibinfo {author} {\bibfnamefont {J.}~\bibnamefont {Shabani}}, \bibinfo {author} {\bibfnamefont {M.}~\bibnamefont {Serbyn}},\ and\ \bibinfo {author} {\bibfnamefont {A.~P.}\ \bibnamefont {Higginbotham}},\ }\bibfield  {title} {\bibinfo {title} {Detecting induced p±ip pairing at the al-{InAs} interface with a quantum microwave circuit},\ }\href {http://dx.doi.org/10.1103/PhysRevLett.128.107701} {\bibfield  {journal} {\bibinfo  {journal} {Phys. Rev. Lett.}\ }\textbf {\bibinfo {volume} {128}},\ \bibinfo {pages} {107701} (\bibinfo {year} {2022})}\BibitemShut {NoStop}%
\bibitem [{\citenamefont {Bøttcher}\ \emph {et~al.}(2024)\citenamefont {Bøttcher}, \citenamefont {Poniatowski}, \citenamefont {Grankin}, \citenamefont {Wesson}, \citenamefont {Yan}, \citenamefont {Vool}, \citenamefont {Galitski},\ and\ \citenamefont {Yacoby}}]{Bottcher2024-kc}%
  \BibitemOpen
  \bibfield  {author} {\bibinfo {author} {\bibfnamefont {C.~G.~L.}\ \bibnamefont {Bøttcher}}, \bibinfo {author} {\bibfnamefont {N.~R.}\ \bibnamefont {Poniatowski}}, \bibinfo {author} {\bibfnamefont {A.}~\bibnamefont {Grankin}}, \bibinfo {author} {\bibfnamefont {M.~E.}\ \bibnamefont {Wesson}}, \bibinfo {author} {\bibfnamefont {Z.}~\bibnamefont {Yan}}, \bibinfo {author} {\bibfnamefont {U.}~\bibnamefont {Vool}}, \bibinfo {author} {\bibfnamefont {V.~M.}\ \bibnamefont {Galitski}},\ and\ \bibinfo {author} {\bibfnamefont {A.}~\bibnamefont {Yacoby}},\ }\bibfield  {title} {\bibinfo {title} {Circuit quantum electrodynamics detection of induced two-fold anisotropic pairing in a hybrid superconductor-ferromagnet bilayer},\ }\href {http://dx.doi.org/10.1038/s41567-024-02613-x} {\bibfield  {journal} {\bibinfo  {journal} {Nat. Phys.}\ }\textbf {\bibinfo {volume} {20}},\ \bibinfo {pages} {1609} (\bibinfo {year} {2024})}\BibitemShut {NoStop}%
\bibitem [{\citenamefont {Hasan}\ and\ \citenamefont {Kane}(2010)}]{Hasan2010-id}%
  \BibitemOpen
  \bibfield  {author} {\bibinfo {author} {\bibfnamefont {M.~Z.}\ \bibnamefont {Hasan}}\ and\ \bibinfo {author} {\bibfnamefont {C.~L.}\ \bibnamefont {Kane}},\ }\bibfield  {title} {\bibinfo {title} {Colloquium: Topological insulators},\ }\href {https://link.aps.org/doi/10.1103/RevModPhys.82.3045} {\bibfield  {journal} {\bibinfo  {journal} {Rev. Mod. Phys.}\ }\textbf {\bibinfo {volume} {82}},\ \bibinfo {pages} {3045} (\bibinfo {year} {2010})}\BibitemShut {NoStop}%
\bibitem [{\citenamefont {Qi}\ and\ \citenamefont {Zhang}(2011)}]{Qi2011-ua}%
  \BibitemOpen
  \bibfield  {author} {\bibinfo {author} {\bibfnamefont {X.-L.}\ \bibnamefont {Qi}}\ and\ \bibinfo {author} {\bibfnamefont {S.-C.}\ \bibnamefont {Zhang}},\ }\bibfield  {title} {\bibinfo {title} {Topological insulators and superconductors},\ }\href {https://link.aps.org/doi/10.1103/RevModPhys.83.1057} {\bibfield  {journal} {\bibinfo  {journal} {Rev. Mod. Phys.}\ }\textbf {\bibinfo {volume} {83}},\ \bibinfo {pages} {1057} (\bibinfo {year} {2011})}\BibitemShut {NoStop}%
\bibitem [{\citenamefont {Fu}\ and\ \citenamefont {Kane}(2008)}]{FuKane2008-bq}%
  \BibitemOpen
  \bibfield  {author} {\bibinfo {author} {\bibfnamefont {L.}~\bibnamefont {Fu}}\ and\ \bibinfo {author} {\bibfnamefont {C.~L.}\ \bibnamefont {Kane}},\ }\bibfield  {title} {\bibinfo {title} {Superconducting proximity effect and majorana fermions at the surface of a topological insulator},\ }\href {https://link.aps.org/doi/10.1103/PhysRevLett.100.096407} {\bibfield  {journal} {\bibinfo  {journal} {Phys. Rev. Lett.}\ }\textbf {\bibinfo {volume} {100}},\ \bibinfo {pages} {096407} (\bibinfo {year} {2008})}\BibitemShut {NoStop}%
\bibitem [{\citenamefont {Schrade}\ and\ \citenamefont {Fu}(2018)}]{Schrade2018-aa}%
  \BibitemOpen
  \bibfield  {author} {\bibinfo {author} {\bibfnamefont {C.}~\bibnamefont {Schrade}}\ and\ \bibinfo {author} {\bibfnamefont {L.}~\bibnamefont {Fu}},\ }\bibfield  {title} {\bibinfo {title} {Majorana superconducting qubit},\ }\href {http://link.aps.org/pdf/10.1103/PhysRevLett.121.267002} {\bibfield  {journal} {\bibinfo  {journal} {Phys. Rev. Lett.}\ }\textbf {\bibinfo {volume} {121}},\ \bibinfo {pages} {267002} (\bibinfo {year} {2018})}\BibitemShut {NoStop}%
\bibitem [{\citenamefont {Hassler}\ \emph {et~al.}(2011)\citenamefont {Hassler}, \citenamefont {Akhmerov},\ and\ \citenamefont {Beenakker}}]{Hassler2011-zr}%
  \BibitemOpen
  \bibfield  {author} {\bibinfo {author} {\bibfnamefont {F.}~\bibnamefont {Hassler}}, \bibinfo {author} {\bibfnamefont {A.~R.}\ \bibnamefont {Akhmerov}},\ and\ \bibinfo {author} {\bibfnamefont {C.~W.~J.}\ \bibnamefont {Beenakker}},\ }\bibfield  {title} {\bibinfo {title} {The top-transmon: a hybrid superconducting qubit for parity-protected quantum computation},\ }\href {http://www.njp.org/} {\bibfield  {journal} {\bibinfo  {journal} {New J. Phys.}\ }\textbf {\bibinfo {volume} {13}},\ \bibinfo {pages} {095004} (\bibinfo {year} {2011})}\BibitemShut {NoStop}%
\bibitem [{\citenamefont {van Heck}\ \emph {et~al.}(2015)\citenamefont {van Heck}, \citenamefont {Hyart},\ and\ \citenamefont {Beenakker}}]{van-Heck2015-en}%
  \BibitemOpen
  \bibfield  {author} {\bibinfo {author} {\bibfnamefont {B.}~\bibnamefont {van Heck}}, \bibinfo {author} {\bibfnamefont {T.}~\bibnamefont {Hyart}},\ and\ \bibinfo {author} {\bibfnamefont {C.~W.~J.}\ \bibnamefont {Beenakker}},\ }\bibfield  {title} {\bibinfo {title} {Minimal circuit for a flux-controlled majorana qubit in a quantum spin-hall insulator},\ }\href {http://dx.doi.org/10.1088/0031-8949/2015/T164/014007} {\bibfield  {journal} {\bibinfo  {journal} {Physica Scripta}\ }\textbf {\bibinfo {volume} {2015}},\ \bibinfo {pages} {014007} (\bibinfo {year} {2015})}\BibitemShut {NoStop}%
\bibitem [{\citenamefont {Ren}\ \emph {et~al.}(2024)\citenamefont {Ren}, \citenamefont {Copenhaver}, \citenamefont {Rokhinson},\ and\ \citenamefont {Väyrynen}}]{Ren2024-ae}%
  \BibitemOpen
  \bibfield  {author} {\bibinfo {author} {\bibfnamefont {Z.}~\bibnamefont {Ren}}, \bibinfo {author} {\bibfnamefont {J.}~\bibnamefont {Copenhaver}}, \bibinfo {author} {\bibfnamefont {L.}~\bibnamefont {Rokhinson}},\ and\ \bibinfo {author} {\bibfnamefont {J.~I.}\ \bibnamefont {Väyrynen}},\ }\bibfield  {title} {\bibinfo {title} {Microwave spectroscopy of majorana vortex modes},\ }\href {https://journals.aps.org/prb/abstract/10.1103/PhysRevB.109.L180506} {\bibfield  {journal} {\bibinfo  {journal} {Phys. Rev. B.}\ }\textbf {\bibinfo {volume} {109}},\ \bibinfo {pages} {L180506} (\bibinfo {year} {2024})}\BibitemShut {NoStop}%
\bibitem [{\citenamefont {Schmitt}\ \emph {et~al.}(2022)\citenamefont {Schmitt}, \citenamefont {Connolly}, \citenamefont {Schleenvoigt}, \citenamefont {Liu}, \citenamefont {Kennedy}, \citenamefont {Chávez-Garcia}, \citenamefont {Jalil}, \citenamefont {Bennemann}, \citenamefont {Trellenkamp}, \citenamefont {Lentz}, \citenamefont {Neumann}, \citenamefont {Lindström}, \citenamefont {de~Graaf}, \citenamefont {Berenschot}, \citenamefont {Tas}, \citenamefont {Mussler}, \citenamefont {Petersson}, \citenamefont {Grützmacher},\ and\ \citenamefont {Schüffelgen}}]{Schmitt2022-il}%
  \BibitemOpen
  \bibfield  {author} {\bibinfo {author} {\bibfnamefont {T.~W.}\ \bibnamefont {Schmitt}}, \bibinfo {author} {\bibfnamefont {M.~R.}\ \bibnamefont {Connolly}}, \bibinfo {author} {\bibfnamefont {M.}~\bibnamefont {Schleenvoigt}}, \bibinfo {author} {\bibfnamefont {C.}~\bibnamefont {Liu}}, \bibinfo {author} {\bibfnamefont {O.}~\bibnamefont {Kennedy}}, \bibinfo {author} {\bibfnamefont {J.~M.}\ \bibnamefont {Chávez-Garcia}}, \bibinfo {author} {\bibfnamefont {A.~R.}\ \bibnamefont {Jalil}}, \bibinfo {author} {\bibfnamefont {B.}~\bibnamefont {Bennemann}}, \bibinfo {author} {\bibfnamefont {S.}~\bibnamefont {Trellenkamp}}, \bibinfo {author} {\bibfnamefont {F.}~\bibnamefont {Lentz}}, \bibinfo {author} {\bibfnamefont {E.}~\bibnamefont {Neumann}}, \bibinfo {author} {\bibfnamefont {T.}~\bibnamefont {Lindström}}, \bibinfo {author} {\bibfnamefont {S.~E.}\ \bibnamefont {de~Graaf}}, \bibinfo {author} {\bibfnamefont {E.}~\bibnamefont {Berenschot}}, \bibinfo {author} {\bibfnamefont {N.}~\bibnamefont {Tas}}, \bibinfo {author}
  {\bibfnamefont {G.}~\bibnamefont {Mussler}}, \bibinfo {author} {\bibfnamefont {K.~D.}\ \bibnamefont {Petersson}}, \bibinfo {author} {\bibfnamefont {D.}~\bibnamefont {Grützmacher}},\ and\ \bibinfo {author} {\bibfnamefont {P.}~\bibnamefont {Schüffelgen}},\ }\bibfield  {title} {\bibinfo {title} {Integration of topological insulator josephson junctions in superconducting qubit circuits},\ }\href {https://pubs.acs.org/doi/10.1021/acs.nanolett.1c04055} {\bibfield  {journal} {\bibinfo  {journal} {Nano Lett.}\ }\textbf {\bibinfo {volume} {22}},\ \bibinfo {pages} {2595} (\bibinfo {year} {2022})}\BibitemShut {NoStop}%
\bibitem [{\citenamefont {Sun}\ \emph {et~al.}(2023)\citenamefont {Sun}, \citenamefont {Li}, \citenamefont {Zhuo}, \citenamefont {Lyu}, \citenamefont {Ji}, \citenamefont {Fan}, \citenamefont {Song}, \citenamefont {Qu}, \citenamefont {Liu}, \citenamefont {Shen},\ and\ \citenamefont {Lu}}]{Sun2023-rj}%
  \BibitemOpen
  \bibfield  {author} {\bibinfo {author} {\bibfnamefont {X.}~\bibnamefont {Sun}}, \bibinfo {author} {\bibfnamefont {B.}~\bibnamefont {Li}}, \bibinfo {author} {\bibfnamefont {E.}~\bibnamefont {Zhuo}}, \bibinfo {author} {\bibfnamefont {Z.}~\bibnamefont {Lyu}}, \bibinfo {author} {\bibfnamefont {Z.}~\bibnamefont {Ji}}, \bibinfo {author} {\bibfnamefont {J.}~\bibnamefont {Fan}}, \bibinfo {author} {\bibfnamefont {X.}~\bibnamefont {Song}}, \bibinfo {author} {\bibfnamefont {F.}~\bibnamefont {Qu}}, \bibinfo {author} {\bibfnamefont {G.}~\bibnamefont {Liu}}, \bibinfo {author} {\bibfnamefont {J.}~\bibnamefont {Shen}},\ and\ \bibinfo {author} {\bibfnamefont {L.}~\bibnamefont {Lu}},\ }\bibfield  {title} {\bibinfo {title} {Realization of superconducting transmon qubits based on topological insulator nanowires},\ }\href {https://pubs.aip.org/aip/apl/article/122/15/154001/2871274} {\bibfield  {journal} {\bibinfo  {journal} {Appl. Phys. Lett.}\ }\textbf {\bibinfo {volume} {122}},\ \bibinfo {pages} {154001} (\bibinfo {year}
  {2023})}\BibitemShut {NoStop}%
\bibitem [{\citenamefont {Koch}\ \emph {et~al.}(2007)\citenamefont {Koch}, \citenamefont {Yu}, \citenamefont {Gambetta}, \citenamefont {Houck}, \citenamefont {Schuster}, \citenamefont {Majer}, \citenamefont {Blais}, \citenamefont {Devoret}, \citenamefont {Girvin},\ and\ \citenamefont {Schoelkopf}}]{Koch2007-vs}%
  \BibitemOpen
  \bibfield  {author} {\bibinfo {author} {\bibfnamefont {J.}~\bibnamefont {Koch}}, \bibinfo {author} {\bibfnamefont {T.~M.}\ \bibnamefont {Yu}}, \bibinfo {author} {\bibfnamefont {J.}~\bibnamefont {Gambetta}}, \bibinfo {author} {\bibfnamefont {A.~A.}\ \bibnamefont {Houck}}, \bibinfo {author} {\bibfnamefont {D.~I.}\ \bibnamefont {Schuster}}, \bibinfo {author} {\bibfnamefont {J.}~\bibnamefont {Majer}}, \bibinfo {author} {\bibfnamefont {A.}~\bibnamefont {Blais}}, \bibinfo {author} {\bibfnamefont {M.~H.}\ \bibnamefont {Devoret}}, \bibinfo {author} {\bibfnamefont {S.~M.}\ \bibnamefont {Girvin}},\ and\ \bibinfo {author} {\bibfnamefont {R.~J.}\ \bibnamefont {Schoelkopf}},\ }\bibfield  {title} {\bibinfo {title} {Charge-insensitive qubit design derived from the cooper pair box},\ }\href {https://journals.aps.org/pra/abstract/10.1103/PhysRevA.76.042319} {\bibfield  {journal} {\bibinfo  {journal} {Phys. Rev. A}\ }\textbf {\bibinfo {volume} {76}},\ \bibinfo {pages} {042319} (\bibinfo {year} {2007})}\BibitemShut {NoStop}%
\bibitem [{\citenamefont {Roth}\ \emph {et~al.}(2023)\citenamefont {Roth}, \citenamefont {Ma},\ and\ \citenamefont {Chew}}]{Roth2023-lq}%
  \BibitemOpen
  \bibfield  {author} {\bibinfo {author} {\bibfnamefont {T.~E.}\ \bibnamefont {Roth}}, \bibinfo {author} {\bibfnamefont {R.}~\bibnamefont {Ma}},\ and\ \bibinfo {author} {\bibfnamefont {W.~C.}\ \bibnamefont {Chew}},\ }\bibfield  {title} {\bibinfo {title} {The transmon qubit for electromagnetics engineers: An introduction},\ }\href {http://dx.doi.org/10.1109/map.2022.3176593} {\bibfield  {journal} {\bibinfo  {journal} {IEEE Antennas Propag. Mag.}\ }\textbf {\bibinfo {volume} {65}},\ \bibinfo {pages} {8} (\bibinfo {year} {2023})}\BibitemShut {NoStop}%
\bibitem [{\citenamefont {Gao}\ \emph {et~al.}(2021)\citenamefont {Gao}, \citenamefont {Rol}, \citenamefont {Touzard},\ and\ \citenamefont {Wang}}]{Gao2021-kj}%
  \BibitemOpen
  \bibfield  {author} {\bibinfo {author} {\bibfnamefont {Y.~Y.}\ \bibnamefont {Gao}}, \bibinfo {author} {\bibfnamefont {M.~A.}\ \bibnamefont {Rol}}, \bibinfo {author} {\bibfnamefont {S.}~\bibnamefont {Touzard}},\ and\ \bibinfo {author} {\bibfnamefont {C.}~\bibnamefont {Wang}},\ }\bibfield  {title} {\bibinfo {title} {Practical guide for building superconducting quantum devices},\ }\href {https://journals.aps.org/prxquantum/abstract/10.1103/PRXQuantum.2.040202} {\bibfield  {journal} {\bibinfo  {journal} {PRX quantum}\ }\textbf {\bibinfo {volume} {2}} (\bibinfo {year} {2021})}\BibitemShut {NoStop}%
\bibitem [{\citenamefont {Xu}\ \emph {et~al.}(2014)\citenamefont {Xu}, \citenamefont {Miotkowski}, \citenamefont {Liu}, \citenamefont {Tian}, \citenamefont {Nam}, \citenamefont {Alidoust}, \citenamefont {Hu}, \citenamefont {Shih}, \citenamefont {Hasan},\ and\ \citenamefont {Chen}}]{Xu2014-fv}%
  \BibitemOpen
  \bibfield  {author} {\bibinfo {author} {\bibfnamefont {Y.}~\bibnamefont {Xu}}, \bibinfo {author} {\bibfnamefont {I.}~\bibnamefont {Miotkowski}}, \bibinfo {author} {\bibfnamefont {C.}~\bibnamefont {Liu}}, \bibinfo {author} {\bibfnamefont {J.}~\bibnamefont {Tian}}, \bibinfo {author} {\bibfnamefont {H.}~\bibnamefont {Nam}}, \bibinfo {author} {\bibfnamefont {N.}~\bibnamefont {Alidoust}}, \bibinfo {author} {\bibfnamefont {J.}~\bibnamefont {Hu}}, \bibinfo {author} {\bibfnamefont {C.-K.}\ \bibnamefont {Shih}}, \bibinfo {author} {\bibfnamefont {M.~Z.}\ \bibnamefont {Hasan}},\ and\ \bibinfo {author} {\bibfnamefont {Y.~P.}\ \bibnamefont {Chen}},\ }\bibfield  {title} {\bibinfo {title} {Observation of topological surface state quantum hall effect in an intrinsic three-dimensional topological insulator},\ }\href {https://idp.nature.com/authorize/casa?redirect_uri=https://www.nature.com/articles/nphys3140&casa_token=J3fX4YWENbUAAAAA:lh5KWowbi7oDH2E2KYnwbyPM8YdjqYm14aoT9_jArvnGIqKln6-CDjP2uuCloPxgJ7J8x76EmPK81At5k3g}
  {\bibfield  {journal} {\bibinfo  {journal} {Nat. Phys.}\ }\textbf {\bibinfo {volume} {10}},\ \bibinfo {pages} {956} (\bibinfo {year} {2014})}\BibitemShut {NoStop}%
\bibitem [{\citenamefont {Jauregui}\ \emph {et~al.}(2018)\citenamefont {Jauregui}, \citenamefont {Kayyalha}, \citenamefont {Kazakov}, \citenamefont {Miotkowski}, \citenamefont {Rokhinson},\ and\ \citenamefont {Chen}}]{Jauregui2018-fr}%
  \BibitemOpen
  \bibfield  {author} {\bibinfo {author} {\bibfnamefont {L.~A.}\ \bibnamefont {Jauregui}}, \bibinfo {author} {\bibfnamefont {M.}~\bibnamefont {Kayyalha}}, \bibinfo {author} {\bibfnamefont {A.}~\bibnamefont {Kazakov}}, \bibinfo {author} {\bibfnamefont {I.}~\bibnamefont {Miotkowski}}, \bibinfo {author} {\bibfnamefont {L.~P.}\ \bibnamefont {Rokhinson}},\ and\ \bibinfo {author} {\bibfnamefont {Y.~P.}\ \bibnamefont {Chen}},\ }\bibfield  {title} {\bibinfo {title} {Gate-tunable supercurrent and multiple andreev reflections in a superconductor-topological insulator nanoribbon-superconductor hybrid device},\ }\href {http://dx.doi.org/10.1063/1.5008746} {\bibfield  {journal} {\bibinfo  {journal} {Appl. Phys. Lett.}\ }\textbf {\bibinfo {volume} {112}},\ \bibinfo {pages} {093105} (\bibinfo {year} {2018})}\BibitemShut {NoStop}%
\bibitem [{\citenamefont {Kayyalha}\ \emph {et~al.}(2019)\citenamefont {Kayyalha}, \citenamefont {Kargarian}, \citenamefont {Kazakov}, \citenamefont {Miotkowski}, \citenamefont {Galitski}, \citenamefont {Yakovenko}, \citenamefont {Rokhinson},\ and\ \citenamefont {Chen}}]{Kayyalha2019-xc}%
  \BibitemOpen
  \bibfield  {author} {\bibinfo {author} {\bibfnamefont {M.}~\bibnamefont {Kayyalha}}, \bibinfo {author} {\bibfnamefont {M.}~\bibnamefont {Kargarian}}, \bibinfo {author} {\bibfnamefont {A.}~\bibnamefont {Kazakov}}, \bibinfo {author} {\bibfnamefont {I.}~\bibnamefont {Miotkowski}}, \bibinfo {author} {\bibfnamefont {V.~M.}\ \bibnamefont {Galitski}}, \bibinfo {author} {\bibfnamefont {V.~M.}\ \bibnamefont {Yakovenko}}, \bibinfo {author} {\bibfnamefont {L.~P.}\ \bibnamefont {Rokhinson}},\ and\ \bibinfo {author} {\bibfnamefont {Y.~P.}\ \bibnamefont {Chen}},\ }\bibfield  {title} {\bibinfo {title} {Anomalous low-temperature enhancement of supercurrent in topological-insulator nanoribbon josephson junctions: Evidence for low-energy andreev bound states},\ }\href {https://link.aps.org/doi/10.1103/PhysRevLett.122.047003} {\bibfield  {journal} {\bibinfo  {journal} {Phys. Rev. Lett.}\ }\textbf {\bibinfo {volume} {122}},\ \bibinfo {pages} {047003} (\bibinfo {year} {2019})}\BibitemShut {NoStop}%
\bibitem [{\citenamefont {Kayyalha}\ \emph {et~al.}(2020)\citenamefont {Kayyalha}, \citenamefont {Kazakov}, \citenamefont {Miotkowski}, \citenamefont {Khlebnikov}, \citenamefont {Rokhinson},\ and\ \citenamefont {Chen}}]{Kayyalha2020-vk}%
  \BibitemOpen
  \bibfield  {author} {\bibinfo {author} {\bibfnamefont {M.}~\bibnamefont {Kayyalha}}, \bibinfo {author} {\bibfnamefont {A.}~\bibnamefont {Kazakov}}, \bibinfo {author} {\bibfnamefont {I.}~\bibnamefont {Miotkowski}}, \bibinfo {author} {\bibfnamefont {S.}~\bibnamefont {Khlebnikov}}, \bibinfo {author} {\bibfnamefont {L.~P.}\ \bibnamefont {Rokhinson}},\ and\ \bibinfo {author} {\bibfnamefont {Y.~P.}\ \bibnamefont {Chen}},\ }\bibfield  {title} {\bibinfo {title} {Highly skewed current–phase relation in superconductor–topological insulator–superconductor josephson junctions},\ }\href {https://doi.org/10.1038/s41535-020-0209-5} {\bibfield  {journal} {\bibinfo  {journal} {npj Quantum Mater.}\ }\textbf {\bibinfo {volume} {5}},\ \bibinfo {pages} {7} (\bibinfo {year} {2020})}\BibitemShut {NoStop}%
\bibitem [{\citenamefont {Jalil}\ \emph {et~al.}(2024)\citenamefont {Jalil}, \citenamefont {Schmitt}, \citenamefont {Rüßmann}, \citenamefont {Wei}, \citenamefont {Frohn}, \citenamefont {Schleenvoigt}, \citenamefont {Wittl}, \citenamefont {Hou}, \citenamefont {Schmidt}, \citenamefont {Underwood}, \citenamefont {Bihlmayer}, \citenamefont {Luysberg}, \citenamefont {Mayer}, \citenamefont {Blügel}, \citenamefont {Grützmacher},\ and\ \citenamefont {Schüffelgen}}]{Jalil2024-cc}%
  \BibitemOpen
  \bibfield  {author} {\bibinfo {author} {\bibfnamefont {A.~R.}\ \bibnamefont {Jalil}}, \bibinfo {author} {\bibfnamefont {T.~W.}\ \bibnamefont {Schmitt}}, \bibinfo {author} {\bibfnamefont {P.}~\bibnamefont {Rüßmann}}, \bibinfo {author} {\bibfnamefont {X.-K.}\ \bibnamefont {Wei}}, \bibinfo {author} {\bibfnamefont {B.}~\bibnamefont {Frohn}}, \bibinfo {author} {\bibfnamefont {M.}~\bibnamefont {Schleenvoigt}}, \bibinfo {author} {\bibfnamefont {W.}~\bibnamefont {Wittl}}, \bibinfo {author} {\bibfnamefont {X.}~\bibnamefont {Hou}}, \bibinfo {author} {\bibfnamefont {A.}~\bibnamefont {Schmidt}}, \bibinfo {author} {\bibfnamefont {K.}~\bibnamefont {Underwood}}, \bibinfo {author} {\bibfnamefont {G.}~\bibnamefont {Bihlmayer}}, \bibinfo {author} {\bibfnamefont {M.}~\bibnamefont {Luysberg}}, \bibinfo {author} {\bibfnamefont {J.}~\bibnamefont {Mayer}}, \bibinfo {author} {\bibfnamefont {S.}~\bibnamefont {Blügel}}, \bibinfo {author} {\bibfnamefont {D.}~\bibnamefont {Grützmacher}},\ and\ \bibinfo {author} {\bibfnamefont
  {P.}~\bibnamefont {Schüffelgen}},\ }\bibfield  {title} {\bibinfo {title} {Engineering epitaxial interfaces for topological insulator — superconductor hybrid devices with al electrodes},\ }\href {https://onlinelibrary.wiley.com/doi/abs/10.1002/qute.202400343} {\bibfield  {journal} {\bibinfo  {journal} {Adv. Quantum Technol.}\ ,\ \bibinfo {pages} {2400343}} (\bibinfo {year} {2024})}\BibitemShut {NoStop}%
\bibitem [{\citenamefont {Paik}\ \emph {et~al.}(2011)\citenamefont {Paik}, \citenamefont {Schuster}, \citenamefont {Bishop}, \citenamefont {Kirchmair}, \citenamefont {Catelani}, \citenamefont {Sears}, \citenamefont {Johnson}, \citenamefont {Reagor}, \citenamefont {Frunzio}, \citenamefont {Glazman}, \citenamefont {Girvin}, \citenamefont {Devoret},\ and\ \citenamefont {Schoelkopf}}]{Paik2011-fm}%
  \BibitemOpen
  \bibfield  {author} {\bibinfo {author} {\bibfnamefont {H.}~\bibnamefont {Paik}}, \bibinfo {author} {\bibfnamefont {D.~I.}\ \bibnamefont {Schuster}}, \bibinfo {author} {\bibfnamefont {L.~S.}\ \bibnamefont {Bishop}}, \bibinfo {author} {\bibfnamefont {G.}~\bibnamefont {Kirchmair}}, \bibinfo {author} {\bibfnamefont {G.}~\bibnamefont {Catelani}}, \bibinfo {author} {\bibfnamefont {A.~P.}\ \bibnamefont {Sears}}, \bibinfo {author} {\bibfnamefont {B.~R.}\ \bibnamefont {Johnson}}, \bibinfo {author} {\bibfnamefont {M.~J.}\ \bibnamefont {Reagor}}, \bibinfo {author} {\bibfnamefont {L.}~\bibnamefont {Frunzio}}, \bibinfo {author} {\bibfnamefont {L.~I.}\ \bibnamefont {Glazman}}, \bibinfo {author} {\bibfnamefont {S.~M.}\ \bibnamefont {Girvin}}, \bibinfo {author} {\bibfnamefont {M.~H.}\ \bibnamefont {Devoret}},\ and\ \bibinfo {author} {\bibfnamefont {R.~J.}\ \bibnamefont {Schoelkopf}},\ }\bibfield  {title} {\bibinfo {title} {Observation of high coherence in josephson junction qubits measured in a three-dimensional circuit
  {QED} architecture},\ }\href {https://journals.aps.org/prl/abstract/10.1103/PhysRevLett.107.240501} {\bibfield  {journal} {\bibinfo  {journal} {Phys. Rev. Lett.}\ }\textbf {\bibinfo {volume} {107}},\ \bibinfo {pages} {240501} (\bibinfo {year} {2011})}\BibitemShut {NoStop}%
\bibitem [{\citenamefont {Krasnok}\ \emph {et~al.}(2024)\citenamefont {Krasnok}, \citenamefont {Dhakal}, \citenamefont {Fedorov}, \citenamefont {Frigola}, \citenamefont {Kelly},\ and\ \citenamefont {Kutsaev}}]{Krasnok2024-xb}%
  \BibitemOpen
  \bibfield  {author} {\bibinfo {author} {\bibfnamefont {A.}~\bibnamefont {Krasnok}}, \bibinfo {author} {\bibfnamefont {P.}~\bibnamefont {Dhakal}}, \bibinfo {author} {\bibfnamefont {A.}~\bibnamefont {Fedorov}}, \bibinfo {author} {\bibfnamefont {P.}~\bibnamefont {Frigola}}, \bibinfo {author} {\bibfnamefont {M.}~\bibnamefont {Kelly}},\ and\ \bibinfo {author} {\bibfnamefont {S.}~\bibnamefont {Kutsaev}},\ }\bibfield  {title} {\bibinfo {title} {Superconducting microwave cavities and qubits for quantum information systems},\ }\href {https://pubs.aip.org/aip/apr/article-pdf/doi/10.1063/5.0155213/18573046/011302_1_5.0155213.pdf} {\bibfield  {journal} {\bibinfo  {journal} {Appl. Phys. Rev.}\ }\textbf {\bibinfo {volume} {11}},\ \bibinfo {pages} {011302} (\bibinfo {year} {2024})}\BibitemShut {NoStop}%
\bibitem [{\citenamefont {Pozar}(2012)}]{Pozar2012-ad}%
  \BibitemOpen
  \bibfield  {author} {\bibinfo {author} {\bibfnamefont {D.~M.}\ \bibnamefont {Pozar}},\ }\href@noop {} {\emph {\bibinfo {title} {Microwave Engineering}}},\ \bibinfo {edition} {4th}\ ed.\ (\bibinfo  {publisher} {John Wiley \& Sons},\ \bibinfo {address} {Nashville, TN},\ \bibinfo {year} {2012})\BibitemShut {NoStop}%
\bibitem [{\citenamefont {Krantz}\ \emph {et~al.}(2019)\citenamefont {Krantz}, \citenamefont {Kjaergaard}, \citenamefont {Yan}, \citenamefont {Orlando}, \citenamefont {Gustavsson},\ and\ \citenamefont {Oliver}}]{Krantz2019-zv}%
  \BibitemOpen
  \bibfield  {author} {\bibinfo {author} {\bibfnamefont {P.}~\bibnamefont {Krantz}}, \bibinfo {author} {\bibfnamefont {M.}~\bibnamefont {Kjaergaard}}, \bibinfo {author} {\bibfnamefont {F.}~\bibnamefont {Yan}}, \bibinfo {author} {\bibfnamefont {T.~P.}\ \bibnamefont {Orlando}}, \bibinfo {author} {\bibfnamefont {S.}~\bibnamefont {Gustavsson}},\ and\ \bibinfo {author} {\bibfnamefont {W.~D.}\ \bibnamefont {Oliver}},\ }\bibfield  {title} {\bibinfo {title} {A quantum engineer's guide to superconducting qubits},\ }\href {https://doi.org/10.1063/1.5089550} {\bibfield  {journal} {\bibinfo  {journal} {Appl. Phys. Rev.}\ }\textbf {\bibinfo {volume} {6}},\ \bibinfo {pages} {021318} (\bibinfo {year} {2019})}\BibitemShut {NoStop}%
\bibitem [{\citenamefont {Pham~Van}\ \emph {et~al.}(1998)\citenamefont {Pham~Van}, \citenamefont {Kurnosikov},\ and\ \citenamefont {Cousty}}]{Pham-Van1998-ey}%
  \BibitemOpen
  \bibfield  {author} {\bibinfo {author} {\bibfnamefont {L.}~\bibnamefont {Pham~Van}}, \bibinfo {author} {\bibfnamefont {O.}~\bibnamefont {Kurnosikov}},\ and\ \bibinfo {author} {\bibfnamefont {J.}~\bibnamefont {Cousty}},\ }\bibfield  {title} {\bibinfo {title} {Evolution of steps on vicinal (0001) surfaces of $\alpha$-alumina},\ }\href {http://dx.doi.org/10.1016/S0039-6028(98)00329-X} {\bibfield  {journal} {\bibinfo  {journal} {Surf. Sci.}\ }\textbf {\bibinfo {volume} {411}},\ \bibinfo {pages} {263} (\bibinfo {year} {1998})}\BibitemShut {NoStop}%
\bibitem [{\citenamefont {Kamal}\ \emph {et~al.}(2016)\citenamefont {Kamal}, \citenamefont {Yoder}, \citenamefont {Yan}, \citenamefont {Gudmundsen}, \citenamefont {Hover}, \citenamefont {Sears}, \citenamefont {Welander}, \citenamefont {Orlando}, \citenamefont {Gustavsson},\ and\ \citenamefont {Oliver}}]{Kamal2016-bf}%
  \BibitemOpen
  \bibfield  {author} {\bibinfo {author} {\bibfnamefont {A.}~\bibnamefont {Kamal}}, \bibinfo {author} {\bibfnamefont {J.}~\bibnamefont {Yoder}}, \bibinfo {author} {\bibfnamefont {F.}~\bibnamefont {Yan}}, \bibinfo {author} {\bibfnamefont {T.}~\bibnamefont {Gudmundsen}}, \bibinfo {author} {\bibfnamefont {D.}~\bibnamefont {Hover}}, \bibinfo {author} {\bibfnamefont {A.}~\bibnamefont {Sears}}, \bibinfo {author} {\bibfnamefont {P.}~\bibnamefont {Welander}}, \bibinfo {author} {\bibfnamefont {T.}~\bibnamefont {Orlando}}, \bibinfo {author} {\bibfnamefont {S.}~\bibnamefont {Gustavsson}},\ and\ \bibinfo {author} {\bibfnamefont {W.}~\bibnamefont {Oliver}},\ }\bibfield  {title} {\bibinfo {title} {Improved superconducting qubit coherence with high-temperature substrate annealing},\ }\href {http://arxiv.org/abs/1606.09262} {\bibfield  {journal} {\bibinfo  {journal} {arXiv: Mesoscale and Nanoscale Physics}\ } (\bibinfo {year} {2016})}\BibitemShut {NoStop}%
\bibitem [{\citenamefont {Son}\ \emph {et~al.}(2020)\citenamefont {Son}, \citenamefont {Shin}, \citenamefont {Zhang}, \citenamefont {Shin}, \citenamefont {Lee}, \citenamefont {Idzuchi}, \citenamefont {Coak}, \citenamefont {Kim}, \citenamefont {Kim}, \citenamefont {Kim}, \citenamefont {Kim}, \citenamefont {Kim}, \citenamefont {Kim},\ and\ \citenamefont {Park}}]{Son2020-kw}%
  \BibitemOpen
  \bibfield  {author} {\bibinfo {author} {\bibfnamefont {S.}~\bibnamefont {Son}}, \bibinfo {author} {\bibfnamefont {Y.~J.}\ \bibnamefont {Shin}}, \bibinfo {author} {\bibfnamefont {K.}~\bibnamefont {Zhang}}, \bibinfo {author} {\bibfnamefont {J.}~\bibnamefont {Shin}}, \bibinfo {author} {\bibfnamefont {S.}~\bibnamefont {Lee}}, \bibinfo {author} {\bibfnamefont {H.}~\bibnamefont {Idzuchi}}, \bibinfo {author} {\bibfnamefont {M.~J.}\ \bibnamefont {Coak}}, \bibinfo {author} {\bibfnamefont {H.}~\bibnamefont {Kim}}, \bibinfo {author} {\bibfnamefont {J.}~\bibnamefont {Kim}}, \bibinfo {author} {\bibfnamefont {J.~H.}\ \bibnamefont {Kim}}, \bibinfo {author} {\bibfnamefont {M.}~\bibnamefont {Kim}}, \bibinfo {author} {\bibfnamefont {D.}~\bibnamefont {Kim}}, \bibinfo {author} {\bibfnamefont {P.}~\bibnamefont {Kim}},\ and\ \bibinfo {author} {\bibfnamefont {J.-G.}\ \bibnamefont {Park}},\ }\bibfield  {title} {\bibinfo {title} {Strongly adhesive dry transfer technique for van der waals heterostructure},\ }\href
  {https://iopscience.iop.org/article/10.1088/2053-1583/abad0b/meta} {\bibfield  {journal} {\bibinfo  {journal} {2d Mater.}\ }\textbf {\bibinfo {volume} {7}},\ \bibinfo {pages} {041005} (\bibinfo {year} {2020})}\BibitemShut {NoStop}%
\bibitem [{\citenamefont {Singh}\ \emph {et~al.}(2014)\citenamefont {Singh}, \citenamefont {Schneider}, \citenamefont {Bosman}, \citenamefont {Merkx},\ and\ \citenamefont {Steele}}]{Singh2014-aw}%
  \BibitemOpen
  \bibfield  {author} {\bibinfo {author} {\bibfnamefont {V.}~\bibnamefont {Singh}}, \bibinfo {author} {\bibfnamefont {B.~H.}\ \bibnamefont {Schneider}}, \bibinfo {author} {\bibfnamefont {S.~J.}\ \bibnamefont {Bosman}}, \bibinfo {author} {\bibfnamefont {E.~P.~J.}\ \bibnamefont {Merkx}},\ and\ \bibinfo {author} {\bibfnamefont {G.~A.}\ \bibnamefont {Steele}},\ }\bibfield  {title} {\bibinfo {title} {Molybdenum-rhenium alloy based high-\textit{Q} superconducting microwave resonators},\ }\href {https://pubs.aip.org/aip/apl/article-pdf/doi/10.1063/1.4903042/12943896/222601_1_online.pdf} {\bibfield  {journal} {\bibinfo  {journal} {Appl. Phys. Lett.}\ }\textbf {\bibinfo {volume} {105}},\ \bibinfo {pages} {222601} (\bibinfo {year} {2014})}\BibitemShut {NoStop}%
\bibitem [{\citenamefont {Huang}\ \emph {et~al.}(2021)\citenamefont {Huang}, \citenamefont {Lienhard}, \citenamefont {Calusine}, \citenamefont {Vepsäläinen}, \citenamefont {Braumüller}, \citenamefont {Kim}, \citenamefont {Melville}, \citenamefont {Niedzielski}, \citenamefont {Yoder}, \citenamefont {Kannan}, \citenamefont {Orlando}, \citenamefont {Gustavsson},\ and\ \citenamefont {Oliver}}]{Huang2021-oz}%
  \BibitemOpen
  \bibfield  {author} {\bibinfo {author} {\bibfnamefont {S.}~\bibnamefont {Huang}}, \bibinfo {author} {\bibfnamefont {B.}~\bibnamefont {Lienhard}}, \bibinfo {author} {\bibfnamefont {G.}~\bibnamefont {Calusine}}, \bibinfo {author} {\bibfnamefont {A.}~\bibnamefont {Vepsäläinen}}, \bibinfo {author} {\bibfnamefont {J.}~\bibnamefont {Braumüller}}, \bibinfo {author} {\bibfnamefont {D.~K.}\ \bibnamefont {Kim}}, \bibinfo {author} {\bibfnamefont {A.~J.}\ \bibnamefont {Melville}}, \bibinfo {author} {\bibfnamefont {B.~M.}\ \bibnamefont {Niedzielski}}, \bibinfo {author} {\bibfnamefont {J.~L.}\ \bibnamefont {Yoder}}, \bibinfo {author} {\bibfnamefont {B.}~\bibnamefont {Kannan}}, \bibinfo {author} {\bibfnamefont {T.~P.}\ \bibnamefont {Orlando}}, \bibinfo {author} {\bibfnamefont {S.}~\bibnamefont {Gustavsson}},\ and\ \bibinfo {author} {\bibfnamefont {W.~D.}\ \bibnamefont {Oliver}},\ }\bibfield  {title} {\bibinfo {title} {Microwave package design for superconducting quantum processors},\ }\href
  {https://journals.aps.org/prxquantum/abstract/10.1103/PRXQuantum.2.020306} {\bibfield  {journal} {\bibinfo  {journal} {PRX quantum}\ }\textbf {\bibinfo {volume} {2}} (\bibinfo {year} {2021})}\BibitemShut {NoStop}%
\bibitem [{\citenamefont {Oostinga}\ \emph {et~al.}(2013)\citenamefont {Oostinga}, \citenamefont {Maier}, \citenamefont {Schüffelgen}, \citenamefont {Knott}, \citenamefont {Ames}, \citenamefont {Brüne}, \citenamefont {Tkachov}, \citenamefont {Buhmann},\ and\ \citenamefont {Molenkamp}}]{Oostinga2013-xi}%
  \BibitemOpen
  \bibfield  {author} {\bibinfo {author} {\bibfnamefont {J.~B.}\ \bibnamefont {Oostinga}}, \bibinfo {author} {\bibfnamefont {L.}~\bibnamefont {Maier}}, \bibinfo {author} {\bibfnamefont {P.}~\bibnamefont {Schüffelgen}}, \bibinfo {author} {\bibfnamefont {D.}~\bibnamefont {Knott}}, \bibinfo {author} {\bibfnamefont {C.}~\bibnamefont {Ames}}, \bibinfo {author} {\bibfnamefont {C.}~\bibnamefont {Brüne}}, \bibinfo {author} {\bibfnamefont {G.}~\bibnamefont {Tkachov}}, \bibinfo {author} {\bibfnamefont {H.}~\bibnamefont {Buhmann}},\ and\ \bibinfo {author} {\bibfnamefont {L.~W.}\ \bibnamefont {Molenkamp}},\ }\bibfield  {title} {\bibinfo {title} {Josephson supercurrent through the topological surface states of strained bulk {HgTe}},\ }\href {http://dx.doi.org/10.1103/PhysRevX.3.021007} {\bibfield  {journal} {\bibinfo  {journal} {Phys. Rev. X.}\ }\textbf {\bibinfo {volume} {3}},\ \bibinfo {pages} {021007} (\bibinfo {year} {2013})}\BibitemShut {NoStop}%
\bibitem [{\citenamefont {Stefanazzi}\ \emph {et~al.}(2022)\citenamefont {Stefanazzi}, \citenamefont {Treptow}, \citenamefont {Wilcer}, \citenamefont {Stoughton}, \citenamefont {Bradford}, \citenamefont {Uemura}, \citenamefont {Zorzetti}, \citenamefont {Montella}, \citenamefont {Cancelo}, \citenamefont {Sussman}, \citenamefont {Houck}, \citenamefont {Saxena}, \citenamefont {Arnaldi}, \citenamefont {Agrawal}, \citenamefont {Zhang}, \citenamefont {Ding},\ and\ \citenamefont {Schuster}}]{Stefanazzi2022-gx}%
  \BibitemOpen
  \bibfield  {author} {\bibinfo {author} {\bibfnamefont {L.}~\bibnamefont {Stefanazzi}}, \bibinfo {author} {\bibfnamefont {K.}~\bibnamefont {Treptow}}, \bibinfo {author} {\bibfnamefont {N.}~\bibnamefont {Wilcer}}, \bibinfo {author} {\bibfnamefont {C.}~\bibnamefont {Stoughton}}, \bibinfo {author} {\bibfnamefont {C.}~\bibnamefont {Bradford}}, \bibinfo {author} {\bibfnamefont {S.}~\bibnamefont {Uemura}}, \bibinfo {author} {\bibfnamefont {S.}~\bibnamefont {Zorzetti}}, \bibinfo {author} {\bibfnamefont {S.}~\bibnamefont {Montella}}, \bibinfo {author} {\bibfnamefont {G.}~\bibnamefont {Cancelo}}, \bibinfo {author} {\bibfnamefont {S.}~\bibnamefont {Sussman}}, \bibinfo {author} {\bibfnamefont {A.}~\bibnamefont {Houck}}, \bibinfo {author} {\bibfnamefont {S.}~\bibnamefont {Saxena}}, \bibinfo {author} {\bibfnamefont {H.}~\bibnamefont {Arnaldi}}, \bibinfo {author} {\bibfnamefont {A.}~\bibnamefont {Agrawal}}, \bibinfo {author} {\bibfnamefont {H.}~\bibnamefont {Zhang}}, \bibinfo {author} {\bibfnamefont {C.}~\bibnamefont
  {Ding}},\ and\ \bibinfo {author} {\bibfnamefont {D.~I.}\ \bibnamefont {Schuster}},\ }\bibfield  {title} {\bibinfo {title} {The {QICK} (quantum instrumentation control kit): Readout and control for qubits and detectors},\ }\href {http://dx.doi.org/10.1063/5.0076249} {\bibfield  {journal} {\bibinfo  {journal} {Rev. Sci. Instrum.}\ }\textbf {\bibinfo {volume} {93}},\ \bibinfo {pages} {044709} (\bibinfo {year} {2022})}\BibitemShut {NoStop}%
\bibitem [{\citenamefont {Ding}\ \emph {et~al.}(2024)\citenamefont {Ding}, \citenamefont {Di~Federico}, \citenamefont {Hatridge}, \citenamefont {Houck}, \citenamefont {Leger}, \citenamefont {Martinez}, \citenamefont {Miao}, \citenamefont {{I, David Schuster}}, \citenamefont {Stefanazzi}, \citenamefont {Stoughton}, \citenamefont {Sussman}, \citenamefont {Treptow}, \citenamefont {Uemura}, \citenamefont {Wilcer}, \citenamefont {Zhang}, \citenamefont {Zhou},\ and\ \citenamefont {Cancelo}}]{Ding2024-vz}%
  \BibitemOpen
  \bibfield  {author} {\bibinfo {author} {\bibfnamefont {C.}~\bibnamefont {Ding}}, \bibinfo {author} {\bibfnamefont {M.}~\bibnamefont {Di~Federico}}, \bibinfo {author} {\bibfnamefont {M.}~\bibnamefont {Hatridge}}, \bibinfo {author} {\bibfnamefont {A.}~\bibnamefont {Houck}}, \bibinfo {author} {\bibfnamefont {S.}~\bibnamefont {Leger}}, \bibinfo {author} {\bibfnamefont {J.}~\bibnamefont {Martinez}}, \bibinfo {author} {\bibfnamefont {C.}~\bibnamefont {Miao}}, \bibinfo {author} {\bibnamefont {{I, David Schuster}}}, \bibinfo {author} {\bibfnamefont {L.}~\bibnamefont {Stefanazzi}}, \bibinfo {author} {\bibfnamefont {C.}~\bibnamefont {Stoughton}}, \bibinfo {author} {\bibfnamefont {S.}~\bibnamefont {Sussman}}, \bibinfo {author} {\bibfnamefont {K.}~\bibnamefont {Treptow}}, \bibinfo {author} {\bibfnamefont {S.}~\bibnamefont {Uemura}}, \bibinfo {author} {\bibfnamefont {N.}~\bibnamefont {Wilcer}}, \bibinfo {author} {\bibfnamefont {H.}~\bibnamefont {Zhang}}, \bibinfo {author} {\bibfnamefont {C.}~\bibnamefont {Zhou}},\ and\
  \bibinfo {author} {\bibfnamefont {G.}~\bibnamefont {Cancelo}},\ }\bibfield  {title} {\bibinfo {title} {Experimental advances with the {QICK} (quantum instrumentation control kit) for superconducting quantum hardware},\ }\href {https://link.aps.org/doi/10.1103/PhysRevResearch.6.013305} {\bibfield  {journal} {\bibinfo  {journal} {Phys. Rev. Res.}\ }\textbf {\bibinfo {volume} {6}},\ \bibinfo {pages} {013305} (\bibinfo {year} {2024})}\BibitemShut {NoStop}%
\bibitem [{\citenamefont {{AMD University Program}}()}]{AMD-RFSoC4x2}%
  \BibitemOpen
  \bibfield  {author} {\bibinfo {author} {\bibnamefont {{AMD University Program}}},\ }\href@noop {} {\bibinfo {title} {{RFSoC 4x2 Kit}}},\ \bibinfo {howpublished} {\url{https://www.amd.com/en/corporate/university-program/aup-boards/rfsoc4x2.html}}\BibitemShut {NoStop}%
\bibitem [{\citenamefont {McRae}\ \emph {et~al.}(2020)\citenamefont {McRae}, \citenamefont {Wang}, \citenamefont {Gao}, \citenamefont {Vissers}, \citenamefont {Brecht}, \citenamefont {Dunsworth}, \citenamefont {Pappas},\ and\ \citenamefont {Mutus}}]{McRae2020-ak}%
  \BibitemOpen
  \bibfield  {author} {\bibinfo {author} {\bibfnamefont {C.~R.~H.}\ \bibnamefont {McRae}}, \bibinfo {author} {\bibfnamefont {H.}~\bibnamefont {Wang}}, \bibinfo {author} {\bibfnamefont {J.}~\bibnamefont {Gao}}, \bibinfo {author} {\bibfnamefont {M.~R.}\ \bibnamefont {Vissers}}, \bibinfo {author} {\bibfnamefont {T.}~\bibnamefont {Brecht}}, \bibinfo {author} {\bibfnamefont {A.}~\bibnamefont {Dunsworth}}, \bibinfo {author} {\bibfnamefont {D.~P.}\ \bibnamefont {Pappas}},\ and\ \bibinfo {author} {\bibfnamefont {J.}~\bibnamefont {Mutus}},\ }\bibfield  {title} {\bibinfo {title} {Materials loss measurements using superconducting microwave resonators},\ }\href {https://doi.org/10.1063/5.0017378} {\bibfield  {journal} {\bibinfo  {journal} {Rev. Sci. Instrum.}\ }\textbf {\bibinfo {volume} {91}},\ \bibinfo {pages} {091101} (\bibinfo {year} {2020})}\BibitemShut {NoStop}%
\bibitem [{scr(2023)}]{scresonators}%
  \BibitemOpen
  \href@noop {} {\bibinfo {title} {{boulder-cryogenic-quantum-testbed/scresonators}}},\ \bibinfo {howpublished} {\url{https://github.com/Boulder-Cryogenic-Quantum-Testbed/scresonators}} (\bibinfo {year} {2023})\BibitemShut {NoStop}%
\bibitem [{\citenamefont {Wang}\ \emph {et~al.}(2021)\citenamefont {Wang}, \citenamefont {Singh}, \citenamefont {McRae}, \citenamefont {Bardin}, \citenamefont {Lin}, \citenamefont {Messaoudi}, \citenamefont {Castelli}, \citenamefont {Rosen}, \citenamefont {Holland}, \citenamefont {Pappas},\ and\ \citenamefont {Mutus}}]{Wang2021-lv}%
  \BibitemOpen
  \bibfield  {author} {\bibinfo {author} {\bibfnamefont {H.}~\bibnamefont {Wang}}, \bibinfo {author} {\bibfnamefont {S.}~\bibnamefont {Singh}}, \bibinfo {author} {\bibfnamefont {C.~R.~H.}\ \bibnamefont {McRae}}, \bibinfo {author} {\bibfnamefont {J.~C.}\ \bibnamefont {Bardin}}, \bibinfo {author} {\bibfnamefont {S.-X.}\ \bibnamefont {Lin}}, \bibinfo {author} {\bibfnamefont {N.}~\bibnamefont {Messaoudi}}, \bibinfo {author} {\bibfnamefont {A.~R.}\ \bibnamefont {Castelli}}, \bibinfo {author} {\bibfnamefont {Y.~J.}\ \bibnamefont {Rosen}}, \bibinfo {author} {\bibfnamefont {E.~T.}\ \bibnamefont {Holland}}, \bibinfo {author} {\bibfnamefont {D.~P.}\ \bibnamefont {Pappas}},\ and\ \bibinfo {author} {\bibfnamefont {J.~Y.}\ \bibnamefont {Mutus}},\ }\bibfield  {title} {\bibinfo {title} {Cryogenic single-port calibration for superconducting microwave resonator measurements},\ }\href {http://dx.doi.org/10.1088/2058-9565/ac070e} {\bibfield  {journal} {\bibinfo  {journal} {Quantum Sci. Technol.}\ }\textbf {\bibinfo {volume}
  {6}},\ \bibinfo {pages} {035015} (\bibinfo {year} {2021})}\BibitemShut {NoStop}%
\bibitem [{\citenamefont {Khalil}\ \emph {et~al.}(2012)\citenamefont {Khalil}, \citenamefont {Stoutimore}, \citenamefont {Wellstood},\ and\ \citenamefont {Osborn}}]{Khalil2012-yi}%
  \BibitemOpen
  \bibfield  {author} {\bibinfo {author} {\bibfnamefont {M.~S.}\ \bibnamefont {Khalil}}, \bibinfo {author} {\bibfnamefont {M.~J.~A.}\ \bibnamefont {Stoutimore}}, \bibinfo {author} {\bibfnamefont {F.~C.}\ \bibnamefont {Wellstood}},\ and\ \bibinfo {author} {\bibfnamefont {K.~D.}\ \bibnamefont {Osborn}},\ }\bibfield  {title} {\bibinfo {title} {An analysis method for asymmetric resonator transmission applied to superconducting devices},\ }\href {http://dx.doi.org/10.1063/1.3692073} {\bibfield  {journal} {\bibinfo  {journal} {J. Appl. Phys.}\ }\textbf {\bibinfo {volume} {111}},\ \bibinfo {pages} {054510} (\bibinfo {year} {2012})}\BibitemShut {NoStop}%
\bibitem [{\citenamefont {Minev}\ \emph {et~al.}(2021)\citenamefont {Minev}, \citenamefont {Leghtas}, \citenamefont {Mundhada}, \citenamefont {Christakis}, \citenamefont {Pop},\ and\ \citenamefont {Devoret}}]{Minev2021-dr}%
  \BibitemOpen
  \bibfield  {author} {\bibinfo {author} {\bibfnamefont {Z.~K.}\ \bibnamefont {Minev}}, \bibinfo {author} {\bibfnamefont {Z.}~\bibnamefont {Leghtas}}, \bibinfo {author} {\bibfnamefont {S.~O.}\ \bibnamefont {Mundhada}}, \bibinfo {author} {\bibfnamefont {L.}~\bibnamefont {Christakis}}, \bibinfo {author} {\bibfnamefont {I.~M.}\ \bibnamefont {Pop}},\ and\ \bibinfo {author} {\bibfnamefont {M.~H.}\ \bibnamefont {Devoret}},\ }\bibfield  {title} {\bibinfo {title} {Energy-participation quantization of josephson circuits},\ }\href {http://dx.doi.org/10.1038/s41534-021-00461-8} {\bibfield  {journal} {\bibinfo  {journal} {Npj Quantum Inf.}\ }\textbf {\bibinfo {volume} {7}},\ \bibinfo {pages} {1} (\bibinfo {year} {2021})}\BibitemShut {NoStop}%
\bibitem [{\citenamefont {Xia}\ \emph {et~al.}(2024)\citenamefont {Xia}, \citenamefont {Huo}, \citenamefont {Li}, \citenamefont {Ying}, \citenamefont {Liu}, \citenamefont {Tang}, \citenamefont {Wang}, \citenamefont {Chen}, \citenamefont {Pan}, \citenamefont {Zhang}, \citenamefont {Liu}, \citenamefont {Li}, \citenamefont {Li}, \citenamefont {He}, \citenamefont {Zhao}, \citenamefont {Shang},\ and\ \citenamefont {Zhang}}]{Xia2024-dz}%
  \BibitemOpen
  \bibfield  {author} {\bibinfo {author} {\bibfnamefont {Z.}~\bibnamefont {Xia}}, \bibinfo {author} {\bibfnamefont {J.}~\bibnamefont {Huo}}, \bibinfo {author} {\bibfnamefont {Z.}~\bibnamefont {Li}}, \bibinfo {author} {\bibfnamefont {J.}~\bibnamefont {Ying}}, \bibinfo {author} {\bibfnamefont {Y.}~\bibnamefont {Liu}}, \bibinfo {author} {\bibfnamefont {X.-Y.}\ \bibnamefont {Tang}}, \bibinfo {author} {\bibfnamefont {Y.}~\bibnamefont {Wang}}, \bibinfo {author} {\bibfnamefont {M.}~\bibnamefont {Chen}}, \bibinfo {author} {\bibfnamefont {D.}~\bibnamefont {Pan}}, \bibinfo {author} {\bibfnamefont {S.}~\bibnamefont {Zhang}}, \bibinfo {author} {\bibfnamefont {Q.}~\bibnamefont {Liu}}, \bibinfo {author} {\bibfnamefont {T.}~\bibnamefont {Li}}, \bibinfo {author} {\bibfnamefont {L.}~\bibnamefont {Li}}, \bibinfo {author} {\bibfnamefont {K.}~\bibnamefont {He}}, \bibinfo {author} {\bibfnamefont {J.}~\bibnamefont {Zhao}}, \bibinfo {author} {\bibfnamefont {R.}~\bibnamefont {Shang}},\ and\ \bibinfo {author} {\bibfnamefont
  {H.}~\bibnamefont {Zhang}},\ }\bibfield  {title} {\bibinfo {title} {Gate-compatible circuit quantum electrodynamics in a three-dimensional cavity architecture},\ }\href {http://dx.doi.org/10.1103/PhysRevApplied.21.034031} {\bibfield  {journal} {\bibinfo  {journal} {Phys. Rev. Appl.}\ }\textbf {\bibinfo {volume} {21}},\ \bibinfo {pages} {034031} (\bibinfo {year} {2024})}\BibitemShut {NoStop}%
\bibitem [{\citenamefont {Krause}\ \emph {et~al.}(2022)\citenamefont {Krause}, \citenamefont {Dickel}, \citenamefont {Vaal}, \citenamefont {Vielmetter}, \citenamefont {Feng}, \citenamefont {Bounds}, \citenamefont {Catelani}, \citenamefont {Fink},\ and\ \citenamefont {Ando}}]{Krause2022-ap}%
  \BibitemOpen
  \bibfield  {author} {\bibinfo {author} {\bibfnamefont {J.}~\bibnamefont {Krause}}, \bibinfo {author} {\bibfnamefont {C.}~\bibnamefont {Dickel}}, \bibinfo {author} {\bibfnamefont {E.}~\bibnamefont {Vaal}}, \bibinfo {author} {\bibfnamefont {M.}~\bibnamefont {Vielmetter}}, \bibinfo {author} {\bibfnamefont {J.}~\bibnamefont {Feng}}, \bibinfo {author} {\bibfnamefont {R.}~\bibnamefont {Bounds}}, \bibinfo {author} {\bibfnamefont {G.}~\bibnamefont {Catelani}}, \bibinfo {author} {\bibfnamefont {J.~M.}\ \bibnamefont {Fink}},\ and\ \bibinfo {author} {\bibfnamefont {Y.}~\bibnamefont {Ando}},\ }\bibfield  {title} {\bibinfo {title} {Magnetic field resilience of three-dimensional transmons with thin-film al/{AlOx}/al josephson junctions approaching 1 {T}},\ }\href {https://journals.aps.org/prapplied/abstract/10.1103/PhysRevApplied.17.034032} {\bibfield  {journal} {\bibinfo  {journal} {Phys. Rev. Appl.}\ }\textbf {\bibinfo {volume} {17}} (\bibinfo {year} {2022})}\BibitemShut {NoStop}%
\bibitem [{\citenamefont {Krause}\ \emph {et~al.}(2024)\citenamefont {Krause}, \citenamefont {Marchegiani}, \citenamefont {Janssen}, \citenamefont {Catelani}, \citenamefont {Ando},\ and\ \citenamefont {Dickel}}]{Krause2024-xg}%
  \BibitemOpen
  \bibfield  {author} {\bibinfo {author} {\bibfnamefont {J.}~\bibnamefont {Krause}}, \bibinfo {author} {\bibfnamefont {G.}~\bibnamefont {Marchegiani}}, \bibinfo {author} {\bibfnamefont {L.~M.}\ \bibnamefont {Janssen}}, \bibinfo {author} {\bibfnamefont {G.}~\bibnamefont {Catelani}}, \bibinfo {author} {\bibfnamefont {Y.}~\bibnamefont {Ando}},\ and\ \bibinfo {author} {\bibfnamefont {C.}~\bibnamefont {Dickel}},\ }\bibfield  {title} {\bibinfo {title} {Quasiparticle effects in magnetic-field-resilient three-dimensional transmons},\ }\href {https://journals.aps.org/prapplied/abstract/10.1103/PhysRevApplied.22.044063} {\bibfield  {journal} {\bibinfo  {journal} {Phys. Rev. Appl.}\ }\textbf {\bibinfo {volume} {22}} (\bibinfo {year} {2024})}\BibitemShut {NoStop}%
\bibitem [{\citenamefont {Larsen}\ \emph {et~al.}(2020)\citenamefont {Larsen}, \citenamefont {Gershenson}, \citenamefont {Casparis}, \citenamefont {Kringhøj}, \citenamefont {Pearson}, \citenamefont {McNeil}, \citenamefont {Kuemmeth}, \citenamefont {Krogstrup}, \citenamefont {Petersson},\ and\ \citenamefont {Marcus}}]{Larsen2020-fi}%
  \BibitemOpen
  \bibfield  {author} {\bibinfo {author} {\bibfnamefont {T.~W.}\ \bibnamefont {Larsen}}, \bibinfo {author} {\bibfnamefont {M.~E.}\ \bibnamefont {Gershenson}}, \bibinfo {author} {\bibfnamefont {L.}~\bibnamefont {Casparis}}, \bibinfo {author} {\bibfnamefont {A.}~\bibnamefont {Kringhøj}}, \bibinfo {author} {\bibfnamefont {N.~J.}\ \bibnamefont {Pearson}}, \bibinfo {author} {\bibfnamefont {R.~P.~G.}\ \bibnamefont {McNeil}}, \bibinfo {author} {\bibfnamefont {F.}~\bibnamefont {Kuemmeth}}, \bibinfo {author} {\bibfnamefont {P.}~\bibnamefont {Krogstrup}}, \bibinfo {author} {\bibfnamefont {K.~D.}\ \bibnamefont {Petersson}},\ and\ \bibinfo {author} {\bibfnamefont {C.~M.}\ \bibnamefont {Marcus}},\ }\bibfield  {title} {\bibinfo {title} {Parity-protected superconductor-semiconductor qubit},\ }\href {https://link.aps.org/doi/10.1103/PhysRevLett.125.056801} {\bibfield  {journal} {\bibinfo  {journal} {Phys. Rev. Lett.}\ }\textbf {\bibinfo {volume} {125}},\ \bibinfo {pages} {056801} (\bibinfo {year} {2020})}\BibitemShut
  {NoStop}%
\bibitem [{\citenamefont {Guo}\ \emph {et~al.}(2024)\citenamefont {Guo}, \citenamefont {Leng},\ and\ \citenamefont {Liu}}]{Guo2024-mo}%
  \BibitemOpen
  \bibfield  {author} {\bibinfo {author} {\bibfnamefont {G.-L.}\ \bibnamefont {Guo}}, \bibinfo {author} {\bibfnamefont {H.-B.}\ \bibnamefont {Leng}},\ and\ \bibinfo {author} {\bibfnamefont {X.}~\bibnamefont {Liu}},\ }\bibfield  {title} {\bibinfo {title} {Parity-spin superconducting qubit based on topological insulators},\ }\href {http://dx.doi.org/10.1088/1367-2630/ad4b58} {\bibfield  {journal} {\bibinfo  {journal} {New J. Phys.}\ }\textbf {\bibinfo {volume} {26}},\ \bibinfo {pages} {063005} (\bibinfo {year} {2024})}\BibitemShut {NoStop}%
\bibitem [{\citenamefont {Douçot}\ and\ \citenamefont {Ioffe}(2012)}]{Doucot2012-ax}%
  \BibitemOpen
  \bibfield  {author} {\bibinfo {author} {\bibfnamefont {B.}~\bibnamefont {Douçot}}\ and\ \bibinfo {author} {\bibfnamefont {L.~B.}\ \bibnamefont {Ioffe}},\ }\bibfield  {title} {\bibinfo {title} {Physical implementation of protected qubits},\ }\href {https://iopscience.iop.org/article/10.1088/0034-4885/75/7/072001} {\bibfield  {journal} {\bibinfo  {journal} {Rep. Prog. Phys.}\ }\textbf {\bibinfo {volume} {75}},\ \bibinfo {pages} {072001} (\bibinfo {year} {2012})}\BibitemShut {NoStop}%
\bibitem [{\citenamefont {Gyenis}\ \emph {et~al.}(2021)\citenamefont {Gyenis}, \citenamefont {Di~Paolo}, \citenamefont {Koch}, \citenamefont {Blais}, \citenamefont {Houck},\ and\ \citenamefont {Schuster}}]{Gyenis2021-vd}%
  \BibitemOpen
  \bibfield  {author} {\bibinfo {author} {\bibfnamefont {A.}~\bibnamefont {Gyenis}}, \bibinfo {author} {\bibfnamefont {A.}~\bibnamefont {Di~Paolo}}, \bibinfo {author} {\bibfnamefont {J.}~\bibnamefont {Koch}}, \bibinfo {author} {\bibfnamefont {A.}~\bibnamefont {Blais}}, \bibinfo {author} {\bibfnamefont {A.~A.}\ \bibnamefont {Houck}},\ and\ \bibinfo {author} {\bibfnamefont {D.~I.}\ \bibnamefont {Schuster}},\ }\bibfield  {title} {\bibinfo {title} {Moving beyond the transmon: Noise-protected superconducting quantum circuits},\ }\href {https://link.aps.org/doi/10.1103/PRXQuantum.2.030101} {\bibfield  {journal} {\bibinfo  {journal} {PRX quantum}\ }\textbf {\bibinfo {volume} {2}},\ \bibinfo {pages} {030101} (\bibinfo {year} {2021})}\BibitemShut {NoStop}%
\bibitem [{\citenamefont {Hegde}\ \emph {et~al.}(2020)\citenamefont {Hegde}, \citenamefont {Yue}, \citenamefont {Wang}, \citenamefont {Huemiller}, \citenamefont {Van~Harlingen},\ and\ \citenamefont {Vishveshwara}}]{Hegde2020-nw}%
  \BibitemOpen
  \bibfield  {author} {\bibinfo {author} {\bibfnamefont {S.~S.}\ \bibnamefont {Hegde}}, \bibinfo {author} {\bibfnamefont {G.}~\bibnamefont {Yue}}, \bibinfo {author} {\bibfnamefont {Y.}~\bibnamefont {Wang}}, \bibinfo {author} {\bibfnamefont {E.}~\bibnamefont {Huemiller}}, \bibinfo {author} {\bibfnamefont {D.~J.}\ \bibnamefont {Van~Harlingen}},\ and\ \bibinfo {author} {\bibfnamefont {S.}~\bibnamefont {Vishveshwara}},\ }\bibfield  {title} {\bibinfo {title} {A topological josephson junction platform for creating, manipulating, and braiding majorana bound states},\ }\href {http://dx.doi.org/10.1016/j.aop.2020.168326} {\bibfield  {journal} {\bibinfo  {journal} {Ann. Phys. (N. Y.)}\ }\textbf {\bibinfo {volume} {423}},\ \bibinfo {pages} {168326} (\bibinfo {year} {2020})}\BibitemShut {NoStop}%
\bibitem [{\citenamefont {Yue}\ \emph {et~al.}(2024)\citenamefont {Yue}, \citenamefont {Zhang}, \citenamefont {Huemiller}, \citenamefont {Montone}, \citenamefont {Arias}, \citenamefont {Wild}, \citenamefont {Zhang}, \citenamefont {Hamilton}, \citenamefont {Yuan}, \citenamefont {Yao}, \citenamefont {Jain}, \citenamefont {Moon}, \citenamefont {Salehi}, \citenamefont {Koirala}, \citenamefont {Oh},\ and\ \citenamefont {Van~Harlingen}}]{Yue2024-nx}%
  \BibitemOpen
  \bibfield  {author} {\bibinfo {author} {\bibfnamefont {G.}~\bibnamefont {Yue}}, \bibinfo {author} {\bibfnamefont {C.}~\bibnamefont {Zhang}}, \bibinfo {author} {\bibfnamefont {E.~D.}\ \bibnamefont {Huemiller}}, \bibinfo {author} {\bibfnamefont {J.~H.}\ \bibnamefont {Montone}}, \bibinfo {author} {\bibfnamefont {G.~R.}\ \bibnamefont {Arias}}, \bibinfo {author} {\bibfnamefont {D.~G.}\ \bibnamefont {Wild}}, \bibinfo {author} {\bibfnamefont {J.~Y.}\ \bibnamefont {Zhang}}, \bibinfo {author} {\bibfnamefont {D.~R.}\ \bibnamefont {Hamilton}}, \bibinfo {author} {\bibfnamefont {X.}~\bibnamefont {Yuan}}, \bibinfo {author} {\bibfnamefont {X.}~\bibnamefont {Yao}}, \bibinfo {author} {\bibfnamefont {D.}~\bibnamefont {Jain}}, \bibinfo {author} {\bibfnamefont {J.}~\bibnamefont {Moon}}, \bibinfo {author} {\bibfnamefont {M.}~\bibnamefont {Salehi}}, \bibinfo {author} {\bibfnamefont {N.}~\bibnamefont {Koirala}}, \bibinfo {author} {\bibfnamefont {S.}~\bibnamefont {Oh}},\ and\ \bibinfo {author} {\bibfnamefont {D.~J.}\ \bibnamefont
  {Van~Harlingen}},\ }\bibfield  {title} {\bibinfo {title} {Signatures of majorana bound states in the diffraction patterns of extended superconductor–topological insulator–superconductor josephson junctions},\ }\href {https://journals.aps.org/prb/abstract/10.1103/PhysRevB.109.094511} {\bibfield  {journal} {\bibinfo  {journal} {Phys. Rev. B.}\ }\textbf {\bibinfo {volume} {109}} (\bibinfo {year} {2024})}\BibitemShut {NoStop}%
\end{thebibliography}%

\end{document}